\documentclass[%
 aps,
 prb,%
 amsmath,amssymb,
reprint%
]{revtex4-2}

\usepackage{graphicx}
\usepackage{dcolumn}
\usepackage{bm}
\usepackage{bbm}
\usepackage{hyperref}
\usepackage{comment}
\usepackage{simpler-wick}

\begin{document}

\title{Charge transport limited by nonlocal electron--phonon interaction. II. Numerically exact quantum dynamics in the slow-phonon regime}

\author{Veljko Jankovi\'c}
 \email{veljko.jankovic@ipb.ac.rs}
 \affiliation{Institute of Physics Belgrade, University of Belgrade, Pregrevica 118, 11080 Belgrade, Serbia}

\begin{abstract}
Transport of charge carriers in mechanically soft semiconductors is mainly limited by their interaction with slow intermolecular phonons.
Carrier motion exhibits a crossover from superdiffusive to subdiffusive, producing a distinct low-frequency peak in the dynamical-mobility profile.
These features can be understood within approaches relying on the timescale separation between carrier and phonon dynamics, such as the transient localization scenario (TLS).
However, recovering them from fully quantum dynamics has proved elusive.
Using the hierarchical equations of motion (HEOM)-based approach exposed in a companion paper (\href{http://arxiv.org/abs/2501.05054}{arXiv:2501.05054}), we study carrier transport in the one-dimensional Peierls model near the adiabatic limit.
We find that the TLS approximates HEOM dynamics very well at higher temperatures and for stronger interactions.
Then, the transport is predominantly phonon-assisted, and turns diffusive from the subdiffusive side well before one phonon period.
In contrast, the band current dominates at moderate temperatures and interactions, relevant for transport in realistic materials.
We then conclude that the super-to-subdiffusive crossover is transient, so that the diffusive motion sets in from the superdiffusive side on timescales comparable to the phonon period.
The low-frequency dynamical mobility then additionally exhibits a dip at approximately one phonon frequency, and the zero-frequency peak.
Our findings in this moderate regime show limitations of the TLS, and support the results of the most advanced quantum--classical simulations.
We expect that the qualitative differences between HEOM and TLS dynamics would diminish for a more realistic phonon density of states.
\end{abstract}

\maketitle
\section{Introduction}
\label{Sec:Introduction}
The prospect of applications in optoelectronic devices has been driving fundamental research into charge transport in halide perovskites~\cite{NanoLett.18.8041,AdvMater.30.1800691,ACSEnergyLett.6.2162} and organic semiconductors~\cite{AdvFunctMater.25.1915,ChemRev.116.13279,NatMater.19.491,JChemPhys.152.190902,AccChemRes.55.819}.
At around room temperature, the main factor influencing charge carrier motion through such materials are the slow, large-amplitude relative motions of the underlying-lattice atoms.
The simplest model of the highly anisotropic charge transport through crystalline organic semiconductors assumes that the vibrations modulate hopping amplitudes between molecules arranged along the direction of the highest conduction~\cite{PhysRevLett.96.086601,AdvMater.19.2000}.
Computing the carrier mobility in the resulting Peierls (or Su--Schrieffer--Heeger) model with nonlocal carrier--phonon interaction~\cite{PhysRevLett.96.086601,PhysRevLett.42.1698,PhysRevB.56.4484,PhysRevLett.103.266601} using the Kubo formula~\cite{Kubo-noneq-stat-mech-book,Mahanbook} faces a number of challenges.

Although the carrier--phonon interaction is typically not excessively strong~\cite{PhysRevLett.109.126407}, the slowness of phonons and their large thermal populations suggest that computations have to go beyond single phonon-assisted processes.
While the approaches based on the polaron transformation~\cite{JChemPhys.83.1854,JChemPhys.83.1843,PhysRevB.69.075211,PhysRevB.69.075212,PhysRevB.69.144302,ApplPhysLett.85.1535,JChemPhys.100.2335,JPhysChemB.115.5312} or exact diagonalization~\cite{PhysRevB.83.165203,PhysRevB.84.245204} treat the interaction nonperturbatively, these typically feature damping or broadening parameters extrinsic to the Hamiltonian.
Quantum Monte Carlo (QMC) simulations provide numerically exact results in the imaginary-frequency domain~\cite{PhysRevLett.114.086601,EPL.125.47002,JChemPhys.156.204116,PhysRevApplied.22.L031004}, yet they rely on the generally uncertain numerical analytical continuation to reconstruct the frequency-dependent mobility.
To capture the diffusive carrier motion, computations based on the time-dependent density matrix renormalization group~\cite{WIREsComputMolSci.12.e1614} often introduce additional vibrational modes, whose interaction with the carrier is Holstein-like~\cite{NatCommun.12.4260,PhysRevB.110.035201}.
Disentangling the effects due to the off-diagonal dynamical disorder from the resulting transport dynamics is then highly nontrivial.
The approaches rooted in the theory of open quantum systems~\cite{JChemPhys.132.081101,JChemPhys.147.214102,JChemPhys.151.044105,JChemPhys.161.084118} usually circumvent the Kubo formula, and compute only the dc mobility by tracking the spread of the carrier starting from a computationally convenient initial condition.
While our approach~\cite{JChemPhys.159.094113,PhysRevB.109.214312} based on the hierarchical equations of motion (HEOM)~\cite{JChemPhys.153.020901,JChemPhys.130.234111,PhysRevE.75.031107} can yield finite-temperature real-time correlation functions, its direct application to the model with off-diagonal dynamical disorder is hampered by the difficulties in treating the phonon-assisted current~\cite{PhysRevB.69.075212,PhysRev.150.529,MolPhys.18.49} after the phonons have been integrated out~\cite{JChemPhys.142.174103,JChemPhys.159.094113}.  
 
The timescale separation between carrier and lattice motions can be used to good advantage as a basis for mixed quantum--classical simulations of coupled carrier--phonon dynamics~\cite{PhysChemChemPhys.17.12395,JChemPhys.156.234109,ChemPhysLett.428.446,PhysRevLett.96.086601,JChemPhys.93.1061,PhysChemChemPhys.21.26368,AccChemRes.55.819,PhysRevB.85.245206,PhysRevB.98.235422,JpnJApplPhys.58.110501}.
The real-time propagation can be avoided altogether by starting from the limit of frozen phonons, in which the long-distance transport is inhibited due to the Anderson localization~\cite{PhysRev.109.1492}, and effectively restoring phonon dynamics (and thus nonzero diffusion constant) using the relaxation-time approximation (RTA)~\cite{PhysRevB.83.081202,PhysRevB.86.245201,PhysRevResearch.2.013001}.
The just described transient localization scenario (TLS)~\cite{AdvFunctMater.26.2292} has become the method of choice for practical computations of carrier mobilities in molecular semiconductors~\cite{NatMater.16.998,JPhysChemC.123.6989}.
Furthermore, the phenomenological TLS-based Drude--Anderson model~\cite{PhysRevB.89.235201,AdvFunctMater.26.2292} has been instrumental in explaining the origin of the finite-frequency peak [the so-called displaced Drude peak (DDP)] in the optical absorption of charge carriers in different materials~\cite{ApplPhysLett.105.143302,ApplPhysLett.120.053302,NatMater.22.1361,PhysRevLett.124.196601}.
Nevertheless, the formal appropriateness of the TLS ansatz, even for the simplest one-dimensional model~\cite{PhysRevLett.96.086601,PhysRevB.83.081202}, has not been rigorously assessed because of the lack of reliable quantum-dynamical insights.

An important original motivation behind devising the TLS~\cite{PhysRevB.83.081202} was to avoid the Ehrenfest dynamics~\cite{ChemPhysLett.428.446,PhysRevLett.96.086601}, which does not preserve the equilibrium distribution of a quantum carrier coupled to classical phonons~\cite{JChemPhys.122.094102}.
As a result, the time-dependent diffusion constant steadily increases instead of reaching its long-time limit~\cite{PhysRevB.83.081202}.
A recent careful reconsideration of the Ehrenfest dynamics~\cite{runeson2024}, as well as novel quantum--classical schemes~\cite{JChemPhys.158.104111,JChemPhys.159.094115,PhysChemChemPhys.26.4929,runeson2024}, have suggested that the long-time growth of the diffusion constant following its decrease on intermediate time scales is not necessarily an artifact of the underlying approximations.
The growth indeed finishes by the diffusion-constant saturation, though on very long time scales~\cite{runeson2024}.
Such a pathway from ballistic to diffusive transport is more involved than that predicted by the TLS~\cite{PhysRevB.83.081202,AdvFunctMater.26.2292} or retrieved from the best available imaginary-axis QMC data~\cite{PhysRevLett.114.086601,EPL.125.47002}.
Consequently, the low-frequency optical response exhibits both the DDP and the standard zero-frequency peak, with a finite-frequency dip in between them~\cite{runeson2024}.
While further evidence in favor of such a rich structure can be found in the quantum--classical simulations of Refs.~\onlinecite{PhysRevX.10.021062,PhysRevB.107.064304}, the results of other quantum--classical approaches~\cite{PhysRevB.98.235422,JpnJApplPhys.58.110501} overall support the TLS, which predicts the DDP and the zero-frequency local minimum~\cite{PhysRevB.83.081202,AdvFunctMater.26.2292}.
In the related one-dimensional Holstein model, our HEOM~\cite{JChemPhys.159.094113,PhysRevB.109.214312} and other numerically exact results~\cite{PhysRevB.72.104304,PhysRevB.106.155129,mitric2024dynamicalquantumtypicalitysimple,mitric2024precursorsandersonlocalizationholstein} reveal that optical responses displaying the rich structure similar to that in Refs.~\onlinecite{runeson2024,PhysRevX.10.021062,PhysRevB.107.064304}, yet different from the corresponding TLS predictions~\cite{PhysRevLett.132.266502}, are ubiquitous for moderate interactions and at moderate-to-high temperatures.

In this study, we demonstrate that our HEOM-based methodology developed in a companion paper~\cite{part1} can be successfully applied near the adiabatic limit of the one-dimensional Peierls model, thus unveiling the details of the ballistic-to-diffusive crossover during charge transport limited by slow off-diagonal dynamical disorder.
We find that the TLS is a very good approximation to the HEOM dynamics only at sufficiently high temperatures and for sufficiently strong interactions.
The hallmark of such parameter regimes is the prevalence of the phonon-assisted current over the band current, and the rapid approach to the diffusive regime, which sets in from the subdiffusive side well before a single phonon period.
In contrast, the band current prevails at moderate temperatures and for moderate interactions, representative of room-temperature transport in rubrene~\cite{AdvMater.19.2000}.
Then, we find that the HEOM transport dynamics is qualitatively similar to that we studied in the one-dimensional Holstein model~\cite{JChemPhys.159.094113,PhysRevB.109.214312}, in which there is no phonon-assisted current.
Our HEOM results expose the transient nature of the super-to-subdiffusive crossover, and establish that the diffusive transport sets in from the superdiffusive side on timescales of the order of the phonon period.
The low-frequency optical response displays the above-discussed rich structure, in contrast to both the TLS~\cite{PhysRevB.83.081202,AdvFunctMater.26.2292} and the best available QMC results~\cite{PhysRevLett.114.086601,EPL.125.47002}.

This manuscript is structured as follows.
Section~\ref{Sec:Model-method} presents the model and methods used in this study.
Section~\ref{Sec:Results} analyzes our HEOM results and compares them to the predictions of the TLS, QMC, and quantum--classical approaches.
In Sec.~\ref{Sec:Discussion}, we critically discuss our main findings, while we summarize them in Sec.~\ref{Sec:Conclusion}.

\section{Model and method}
\label{Sec:Model-method}
\subsection{Model}
\label{SSec:Model}
We consider the one-dimensional Peierls model containing a single charge carrier in the field of dispersionless optical phonons that modulate its kinetic energy~\cite{PhysRevLett.96.086601,PhysRevB.83.081202,AdvFunctMater.26.2292}.
In the following, we set the lattice constant $a_l$, the elementary charge $e_0$, and physical constants $\hbar$ and $k_B$ to unity.
The model Hamiltonian reads
\begin{equation}
\label{Eq:H_tot}
    H=H_\mathrm{e}+H_\mathrm{e-ph}+H_\mathrm{ph},
\end{equation}
where
\begin{equation}
\label{Eq:H_e_plus_H_e_ph}
\begin{split}
    H_\mathrm{e}+H_\mathrm{e-ph}=&\sum_n\left[-J+g\left(b_n+b_n^\dagger-b_{n+1}-b_{n+1}^\dagger\right)\right]\times\\&\left(|n\rangle\langle n+1|+|n+1\rangle\langle n|\right)
\end{split}
\end{equation}
governs phonon-modulated nearest-neighbor carrier hops, whereas
\begin{equation}
\label{Eq:H_ph}
    H_\mathrm{ph}=\omega_0\sum_n b_n^\dagger b_n
\end{equation}
is the free-phonon Hamiltonian.
In Eq.~\eqref{Eq:H_e_plus_H_e_ph}, $|n\rangle$ is the single-electron state localized on site $n$, and bosonic operators $b_n$ ($b_n^\dagger$) create (annihilate) a phonon on site $n$.
The characteristic energy scales of the model are the hopping amplitude $J$, the phonon frequency $\omega_0$, and the electron--phonon interaction constant $g$.
These are conveniently combined into two dimensionless ratios, the adiabaticity parameter $\omega_0/J$ and the dimensionless carrier--phonon interaction
\begin{equation}
    \lambda=\frac{2g^2}{\omega_0J}.
\end{equation}

Comprehensive insights into charge transport can be gained from the finite-temperature real-time current--current correlation function
\begin{equation}
\label{Eq:def_C_jj}
    C_{jj}(t)=\langle j(t)j(0)\rangle=\mathrm{Tr}\{e^{iHt}je^{-iHt}j\rho^\mathrm{eq}\}.
\end{equation}
The angular brackets in Eq.~\eqref{Eq:def_C_jj} denote averaging over the equilibrium state $\rho^\mathrm{eq}=e^{-\beta H}/\mathrm{Tr}\:e^{-\beta H}$ of the interacting carrier--phonon system at temperature $T=\beta^{-1}$.
The current operator
\begin{equation}
\label{Eq:def_j}
\begin{split}
    j&=j_\mathrm{e}+j_\mathrm{e-ph}\\&=-i\sum_n\left[-J+g\left(b_n+b_n^\dagger-b_{n+1}-b_{n+1}^\dagger\right)\right]\times\\&\left(|n\rangle\langle n+1|-|n+1\rangle\langle n|\right)
\end{split}
\end{equation}
has the so-called phonon-assisted contribution $j_\mathrm{e-ph}$ [the term proportional to $g$ in Eq.~\eqref{Eq:def_j}] in addition to the band contribution $j_\mathrm{e}$ [the term proportional to $J$ in Eq.~\eqref{Eq:def_j}].
A more intuitive understanding of transport dynamics is offered by the time-dependent diffusion constant
\begin{equation}
    \mathcal{D}(t)=\frac{1}{2}\frac{d}{dt}\Delta x^2(t)=\int_0^t ds\:\mathrm{Re}\:C_{jj}(s),
\end{equation}
which determines the growth rate of the carrier's mean-square displacement
\begin{equation}
    \Delta x^2(t)=\left\langle[x(t)-x(0)]^2\right\rangle,
\end{equation}
with $x$ being the carrier's position operator.
The carrier motion changes its character from the short-time ballistic motion characterized by $\Delta x^2(t)\propto t^2$ and $\mathcal{D}(t)\propto t$ to the long-time diffusive motion, when $\Delta x^2(t)\propto t$ and $\mathcal{D}(t)=\mathcal{D}_\infty$.
The carrier mobility $\mu_\mathrm{dc}$ then follows from the Einstein relation $\mu_\mathrm{dc}=\frac{\mathcal{D}_\infty}{T}$.
Another convenient quantity to track the ballistic-to-diffusive crossover is the diffusion exponent $\alpha(t)\geq 0$, defined by $\Delta x^2(t)\propto t^{\alpha(t)}$ or
\begin{equation}
    \alpha(t)=\frac{2t\mathcal{D}(t)}{\Delta x^2(t)}.
\end{equation}
In realistic systems, the crossover dynamics is often inferred from the experimentally accessible carriers' optical response, which is proportional to the dynamical mobility
\begin{equation}
    \mathrm{Re}\:\mu(\omega)=\frac{1-e^{-\beta\omega}}{2\omega}\int_{-\infty}^{+\infty}dt\:e^{i\omega t}C_{jj}(t).
\end{equation}

\subsection{Hierarchical equations of motion}
The HEOM method offers a well-established numerically exact framework to study the dynamics of a system of interest (here, the carrier) coupled to a collection of harmonic oscillators (here, phonons)~\cite{JChemPhys.153.020901,JChemPhys.130.234111,PhysRevE.75.031107}.
Nevertheless, it has been quite challenging to fit the computations of finite-temperature real-time correlation functions of mixed carrier--phonon operators, such as the current operator in Eq.~\eqref{Eq:def_j}, into the HEOM framework, which straightforwardly treats only purely carrier operators~\cite{PhysRevB.105.054311,JChemPhys.159.094113,PhysRevB.109.214312,JChemPhys.142.174103,JChemPhys.143.194106,JChemPhys.156.244102,PhysRevLett.109.266403,JPhysChemLett.15.1382}.
While the related dissipaton-equations-of-motion (DEOM) approach~\cite{JChemPhys.140.054105,FrontPhys.11.110306,MolPhys.116.780,JChemPhys.157.170901} can, in principle, address the challenge, its direct applications to models that feature a \emph{finite} collection of \emph{undamped} harmonic oscillators, and thus apparently lack dissipation~\cite{JChemPhys.150.184109,JChemPhys.153.204109,JChemPhys.160.111102}, can be problematic.
In such a scenario, our companion paper~\cite{part1} explicitly expresses HEOM auxiliaries in terms of phonon creation and annihilation operators, and formulates a transparent procedure to handle the phonon-assisted current. 
We perform HEOM computations on an $N$-site chain with periodic boundary conditions ($|N+1\rangle\equiv|1\rangle,b_{N+1}\equiv b_1$).
We represent the dynamical equations in momentum space, see Sec.~IV.A of Ref.~\onlinecite{part1}, which reduces their total number, and truncate the hierarchy at a sufficiently large maximum depth $D$.
In models that apparently lack dissipation, the truncated HEOM exhibits numerical instabilities that prevent reliable computations of the long-time carrier dynamics and carrier mobility~\cite{JChemPhys.150.184109,JChemPhys.153.204109,JChemPhys.160.111102}.
We mitigate these instabilities by resorting to the Markovian-adiabatic hierarchy closing scheme we developed in Refs.~\onlinecite{JChemPhys.159.094113,part1}, which stabilizes the long-time dynamics without introducing closing-specific artifacts that compromise HEOM results for $\mu_\mathrm{dc}$.
For more details, consult Refs.~\onlinecite{JChemPhys.159.094113,part1}, as well as Appendix~\ref{App:closing-scheme}.

The current-operator decomposition in Eq.~\eqref{Eq:def_j} implies the decomposition of $\mathcal{D}(t)=\sum_c\mathcal{D}_c(t)$ and $\mathrm{Re}\:\mu(\omega)=\sum_c\mathrm{Re}\:\mu_c(\omega)$ into the purely carrier (band, $c$=e), phonon-assisted ($c$=ph), and cross ($c$=x) contributions.
As discussed in Sec.~V of Ref.~\onlinecite{part1}, the character of the transport is conveniently discussed in terms of the following phonon-assisted $S_\mathrm{ph}\geq 0$ and cross $S_\mathrm{x}\leq 0$ shares of the dc mobility:
\begin{equation}
    S_c=\frac{\mu_\mathrm{dc}^c}{\mu_\mathrm{dc}^\mathrm{e}+\mu_\mathrm{dc}^\mathrm{ph}}.
\end{equation}
Section~SI of the Supplemental Material~\cite{comment261224} summarizes the parameter regimes and numerical parameters of our HEOM computations.
The data that support our conclusions are openly available~\cite{jankovic_2025_14637019}.

\subsection{Transient localization scenario}
The TLS is a widely used physically motivated~\cite{PhysRevB.83.081202,AdvFunctMater.26.2292,PhysRevResearch.2.013001,PhysRevLett.132.266502} and computationally convenient~\cite{NatMater.16.998,JPhysChemC.123.6989,PhysRevLett.124.196601} approach to computing $C_{jj}(t)$ in the limit of slow ($\omega_0/J\ll 1$) and abundantly present ($T\gtrsim\omega_0$) phonons.
Then, charge dynamics on timescales short compared to $\omega_0^{-1}$, when phonons can be considered as frozen, is essentially the same as the dynamics in the presence of Gaussian static disorder in the hopping amplitude, whose strength is $\sigma^2=2\frac{2g^2}{\beta\omega_0}=2\lambda JT$.
Formally, one replaces the phonon operator $g\left(b_n+b_n^\dagger-b_{n+1}-b_{n+1}^\dagger\right)$ in Eqs.~\eqref{Eq:H_e_plus_H_e_ph} and~\eqref{Eq:def_j} by a Gaussian random variable $X_{n,n+1}$ of zero mean and variance $\sigma^2$, and the trace in Eq.~\eqref{Eq:def_C_jj} by the average over different disorder realizations.
The current--current correlation function computed upon introducing these assumptions in Eq.~\eqref{Eq:def_C_jj} is denoted as $C_{jj}^\mathrm{dis}(t)$.
Charge diffusion, which is inhibited in the static-disorder setup~\cite{PhysRev.109.1492}, is ultimately established through the coupled charge--phonon dynamics, whose effects become appreciable on timescale $\tau_d$.
The TLS effectively restores phonon dynamics by virtue of the RTA, in which
\begin{equation}
\label{Eq:TLS_ansatz}
    C_{jj}^\mathrm{TLS}(t)=C_{jj}^\mathrm{dis}(t)e^{-|t|/\tau_d}.
\end{equation}
Physically, $\tau_d^{-1}=\alpha_d\omega_0$, with the proportionality constant $\alpha_d\sim 1$~\cite{PhysRevB.83.081202}.
Its precise value can be determined by requiring that the TLS predictions for transport properties reasonably agree with some reference results emerging from fully quantum charge--phonon dynamics.
Considering a modification of Eq.~\eqref{Eq:TLS_ansatz}, the authors of Ref.~\onlinecite{PhysRevResearch.2.013001} concluded that the best overall agreement between the temperature-dependent mobilities obtained using the QMC~\cite{PhysRevLett.114.086601} and the modified TLS is achieved for $\alpha_d=2.2$.
Studying the related Holstein model, the same authors reported that the same value of $\alpha_d$ brings the TLS results closest to their reference numerically exact results~\cite{PhysRevLett.132.266502}.
Therefore, we adopt $\alpha_d=2.2$ in all our TLS computations, whose details are provided in Sec.~SII of the Supplemental Material~\cite{comment261224}.
While reasonable variations of $\alpha_d$ around unity are known to influence quantitative predictions for $\mu_\mathrm{dc}$~\cite{PhysRevResearch.2.013001}, the overall physical picture offered by the TLS is robust against these variations.     

\section{Results}
\label{Sec:Results}

\begin{figure*}[htbp!]
    \centering
    \includegraphics[width=\textwidth]{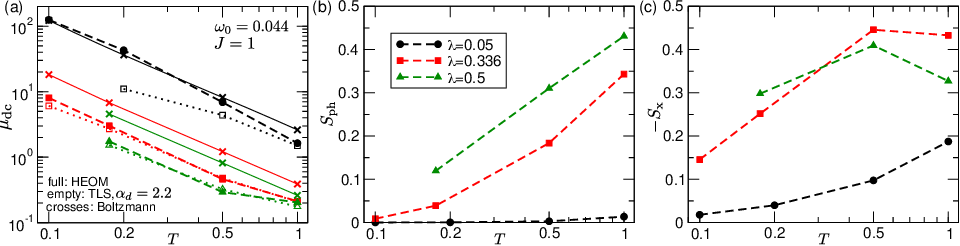}
    \caption{(a) Temperature-dependent carrier mobility $\mu_\mathrm{dc}(T)$ computed using the HEOM method (full symbols connected with dashed lines), the TLS with $\alpha_d=2.2$ in Eq.~\eqref{Eq:TLS_ansatz} (empty symbols connected with dotted lines), and the Boltzmann theory in the momentum relaxation-time approximation (crosses connected with the solid line). (b) and (c): Shares of the phonon-assisted [$S_\mathrm{ph}$, (b)] and cross [$-S_\mathrm{x}$, (c)] contributions to $\mu_\mathrm{dc}(T)$ computed using the HEOM method.}
    \label{Fig:mu_dc_vs_T_w0_0.044_300924}
\end{figure*}

We examine one-dimensional transport of a charge that is weakly to moderately ($\lambda\lesssim 0.5$) coupled to slow ($\omega_0/J\sim 0.05$) quantum phonons, which are abundantly thermally excited ($T/\omega_0\gtrsim 1$).
While we mostly analyze the regime $\omega_0/J=0.044$, $\lambda=0.336$, and $T/J=0.175$ ($T/\omega_0\approx 4$), which is representative of the room-temperature transport along the direction of maximal conduction in single-crystal rubrene~\cite{AdvMater.19.2000,AdvFunctMater.26.2292}, we also study how variations in $T$ and $\lambda$ affect transport properties.
In physical units, we set $J=143\:\mathrm{meV}$ ($\approx 1150\:\mathrm{cm}^{-1}$), $\hbar\omega_0=6.2\:\mathrm{meV}$ ($\approx 50\:\mathrm{cm}^{-1}$), and $a_l=7.2\:\mathrm{\AA}$~\cite{AdvMater.19.2000}, so that the mobility is measured in units of $\mu_\mathrm{dc}^\mathrm{unit}=e_0a_l^2/\hbar=7.9\:\mathrm{cm}^2/(\mathrm{Vs})$.
As discussed in the companion paper~\cite{part1}, our HEOM computations are feasible at temperatures above $T_\mathrm{min}/\omega_0\approx 2$, i.e., $T_\mathrm{min}/J\approx 0.1$ or $T_\mathrm{min}\approx 150\:\mathrm{K}$.
At low temperatures, charge transport is also influenced by other scattering mechanisms (e.g., impurity scattering)~\cite{EPL.125.47002}, rendering the model studied here of limited relevance.

Our main results are summarized in Fig.~\ref{Fig:mu_dc_vs_T_w0_0.044_300924} that shows the temperature-dependent carrier mobility [Fig.~\ref{Fig:mu_dc_vs_T_w0_0.044_300924}(a)] along with the relative importance of its phonon-assisted [Fig.~\ref{Fig:mu_dc_vs_T_w0_0.044_300924}(b)] and cross [Fig.~\ref{Fig:mu_dc_vs_T_w0_0.044_300924}(c)] contributions.
Before all else, although the phonon-assisted contribution to $\mu_\mathrm{dc}$ around room temperature ($T/J\approx 0.2$) is much smaller than the purely electronic contribution, see Fig.~\ref{Fig:mu_dc_vs_T_w0_0.044_300924}(b), the cross contribution to $\mu_\mathrm{dc}$ cannot be neglected, see Fig.~\ref{Fig:mu_dc_vs_T_w0_0.044_300924}(c).
The only exception is the weak-interaction regime ($\lambda=0.05$), in which the predictions of the Boltzmann transport theory closely follow our HEOM mobilities up to very high temperatures, see Fig.~\ref{Fig:mu_dc_vs_T_w0_0.044_300924}(a), at which the cross contribution gains importance.
Therefore, Figs.~\ref{Fig:mu_dc_vs_T_w0_0.044_300924}(b) and~\ref{Fig:mu_dc_vs_T_w0_0.044_300924}(c) show that approximate approaches attempting at accurately evaluating the mobility of a carrier moderately coupled to slow quantum phonons should \emph{a priori} take into account the cross contribution to $\mu_\mathrm{dc}$.
We note that many approximate approaches omit it altogether~\cite{PhysRevLett.103.266601,PhysRev.150.529,JChemPhys.70.3775,PhysRevB.69.075212}.
The details of our computations using the Boltzmann transport theory within the momentum relaxation-time approximation are provided in Sec.~SIII of the Supplemental Material~\cite{comment261224}.

\begin{figure}[htbp!]
    \centering
    \includegraphics[width=\columnwidth]{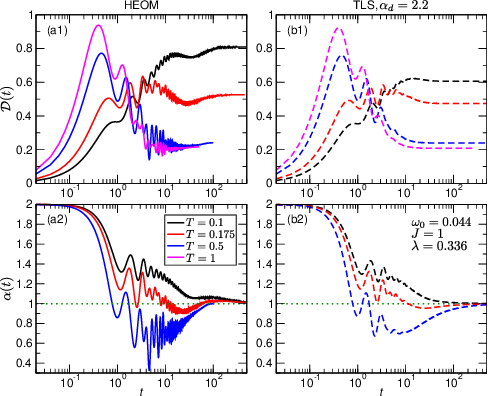}
    \caption{Dynamics of the diffusion constant [(a1) and (b1)] and diffusion exponent [(a2) and (b2)] computed using HEOM [(a1) and (a2)] and TLS with $\alpha_d=2.2$ in Eq.~\eqref{Eq:TLS_ansatz} [(b1) and (b2)] at different temperatures.
    The dotted lines in (a2) and (b2) display the diffusive limit $\alpha(t)=1$.
    The model parameters are set to $J=1,\omega_0=0.044,\lambda=0.336$. 
    The temperature $T=0.175$ is representative of the room-temperature transport.
    For visual clarity, we omit the dynamics of the diffusion exponent at $T=1$ from the lower panels.
    }
    \label{Fig:diffusion_exponents_230924}
\end{figure}

\subsection{Comparison of HEOM and imaginary-axis QMC results}
We first compare our results with the best currently available numerically exact results~\cite{PhysRevLett.114.086601}.
For $\lambda=0.336$ and at $T/J=0.175$ ($T/J=0.1$), we find $\mu_\mathrm{dc}^\mathrm{HEOM}\approx 3$ ($\mu_\mathrm{dc}^\mathrm{HEOM}\approx 8$), which corresponds to $\mu_\mathrm{dc}^\mathrm{HEOM}\approx 23\:\mathrm{cm}^2/(\mathrm{Vs})$ [$\mu_\mathrm{dc}^\mathrm{HEOM}\approx 63\:\mathrm{cm}^2/(\mathrm{Vs})$] in physical units.
These values agree reasonably well with the values reported in Ref.~\onlinecite{PhysRevLett.114.086601}, which amount to $\mu_\mathrm{dc}^\mathrm{QMC}\approx 22\:\mathrm{cm}^2/(\mathrm{Vs})$ [$\mu_\mathrm{dc}^\mathrm{QMC}\approx 50\:\mathrm{cm}^2/(\mathrm{Vs})$] at $T/J=0.175$ ($T/J=0.1$)~\footnote{The authors of Ref.~\onlinecite{PhysRevLett.114.086601} use $J=93\:\mathrm{meV}$, and to enable a direct comparison between our HEOM and their QMC mobilities in physical units, we rescale their results by $143/93$, as was done in Ref.~\onlinecite{AdvFunctMater.26.2292}.}.
Also, $\mu_\mathrm{dc}^\mathrm{HEOM}$ at $T/J=0.175$ falls in the range of the experimentally measured room-temperature mobilities in rubrene [$\sim 10-20\:\mathrm{cm}^2/(\mathrm{Vs})$]~\cite{PhysRevLett.93.086602,PhysRevLett.95.226601,PhysRevLett.98.196804}.

Although the carrier mobility conveniently encodes information on carrier dynamics on all time and length scales, its value can be relatively insensitive to the details of the dynamics~\cite{PhysRevB.109.214312}, which are presented in Figs.~\ref{Fig:diffusion_exponents_230924}(a1) and~\ref{Fig:diffusion_exponents_230924}(a2).
We find that the carrier motion during the ballistic-to-diffusive crossover is superdiffusive at lower temperatures [$\alpha(t)>1$ at $T/J=0.1$], while it displays subdiffusive [$\alpha(t)<1$] features at higher temperatures.
Our results overall agree with Ref.~\onlinecite{PhysRevLett.114.086601}, which establishes that the hallmark of charge dynamics in the field of slow, large-amplitude intermolecular vibrations is the gradual change from super- to subdiffusive dynamics with increasing temperature.
However, in contrast to Ref.~\onlinecite{PhysRevLett.114.086601}, we find that the subdiffusive carrier dynamics at room temperature is temporally limited, see the curve for $T/J=0.175$ in Fig.~\ref{Fig:diffusion_exponents_230924}(a2).
The subdiffusion then extends to approximately one half of the phonon period, after which the diffusive transport ultimately sets in from the superdiffusive side.
Such dynamics of the diffusion exponent reflect the increase of the diffusion constant that starts at $t_\mathrm{min}\approx 1/\omega_0$, when both $\mathcal{D}$ and $\alpha$ reach their respective local minima, and terminates by its saturation on a timescale $t_\mathrm{max}\sim 2\pi/\omega_0$ comparable to the phonon period.
Meanwhile, at a higher temperature $T/J=0.5$, at which the diffusion constant plateaus well before the phonon period (already for $t\gtrsim t_\mathrm{min}$), the diffusive transport sets in from the subdiffusive side, in agreement with the conclusions of Ref.~\onlinecite{PhysRevLett.114.086601}.

The above-discussed qualitative differences between HEOM and QMC transport dynamics deserve additional comments, as both approaches are numerically exact.
The HEOM results for $\alpha(t)$ at $T/J=0.175$ are very close to unity for $Jt\gtrsim 10$, the relative deviation not surpassing 10\%.
In a similar vein, the increase in $\mathcal{D}(t)$ for $t\gtrsim t_\mathrm{min}$ is such that the local minimum $\mathcal{D}(t_\mathrm{min})$ lies within the 10\% errorbar associated with the long-time value $\mathcal{D}_\infty$~\cite{part1}.
One might suspect that the HEOM dynamics of $\alpha$ and $\mathcal{D}$ at $T/J=0.175$ reflect artifacts due to the insufficient chain length $N$ or insufficient maximum depth $D$ or the approximations underlying the Markovian-adiabatic closing scheme, which is necessary to obtain meaningful dynamics on intermediate-to-long timescales~\cite{part1}.
In Appendix~\ref{App:finite-N-and-D}, we check that increasing $N$ or $D$ does not qualitatively change the dynamics of $\alpha$ and $\mathcal{D}$, while quantitative changes are consistent with the errorbars accompanying our HEOM results.
Our discussion in Appendix~\ref{App:closing-scheme} highlights that the details of the approximations underlying the most commonly used closing schemes (see Sec.~IV.C of the companion paper~\cite{part1}) have no qualitative (and almost no quantitative) influence on the dynamics of $\alpha$ and $\mathcal{D}$.
The closing scheme mainly serves to stabilize the dynamics, which displays the increase in $\mathcal{D}(t)$ for $t\gtrsim t_\mathrm{min}$ even when the closing terms are set to zero, i.e., even when the HEOM are merely truncated without closing.
Therefore, the qualitative differences between our HEOM dynamics and those of Ref.~\onlinecite{PhysRevLett.114.086601} at $T/J=0.175$ can be most probably ascribed to (statistical) uncertainties of the procedures for numerical analytical continuation.
The small amplitude of the features in $\alpha(t)$ and $\mathcal{D}(t)$ on intermediate-to-long timescales renders their reconstruction from generally noisy imaginary-axis data rather challenging~\cite{Buividovich:2024MY}.
Notably, this circumstance does not compromise QMC results for the carrier mobility, which, as an integrated quantity, is relatively insensitive to the details of transport dynamics.
This example once again stresses that the details of charge transport can be fully uncovered only using real-axis methods~\cite{PhysRevB.109.214312,PhysRevLett.123.036601,mitric2024precursorsandersonlocalizationholstein}.

The HEOM transport dynamics for $T/J=0.175$ in Figs.~\ref{Fig:diffusion_exponents_230924}(a1) and~\ref{Fig:diffusion_exponents_230924}(a2) and Appendices~\ref{App:finite-N-and-D} and~\ref{App:closing-scheme} might seem somewhat unphysical, as its details are not fully compatible with the intuition based on simplified physically appealing models~\cite{PhysRevB.83.081202,AdvFunctMater.26.2292,PhysRevLett.114.086601}.
We reiterate (see Sec.~\ref{Sec:Introduction}) that temporally limited subdiffusive dynamics~\cite{JChemPhys.159.094113,PhysRevB.109.214312}, the upturn in the diffusion constant on intermediate timescales~\cite{JChemPhys.159.094113,PhysRevB.109.214312,mitric2024precursorsandersonlocalizationholstein,mitric2024dynamicalquantumtypicalitysimple}, and the dynamical-mobility profile displaying both the DDP and the zero-frequency peak (see Sec.~\ref{SSec:HEOM-vs-TLS})~\cite{PhysRevB.72.104304,PhysRevB.109.214312,mitric2024precursorsandersonlocalizationholstein,mitric2024dynamicalquantumtypicalitysimple} have been repeatedly reported in real-axis numerically exact studies of the related Holstein model at moderate interactions and moderate-to-high temperatures.
Moreover, qualitatively similar dynamics emerge from the most recent quantum--classical simulations of the Peierls model~\cite{runeson2024}.
The common feature of all these studies is that they take seriously the dynamics of undamped phonon modes in Eqs.~\eqref{Eq:H_tot}--\eqref{Eq:H_ph}, either numerically exactly~\cite{PhysRevB.72.104304,JChemPhys.159.094113,PhysRevB.109.214312,mitric2024precursorsandersonlocalizationholstein,mitric2024dynamicalquantumtypicalitysimple} or in appropriate approximations~\cite{mitric2024precursorsandersonlocalizationholstein,runeson2024}.
Meanwhile, the TLS ansatz or the picture of phonons as a viscous medium that slows the carrier down~\cite{PhysRevLett.114.086601} introduce effective phonon dynamics, and thus effectively change the model Hamiltonian.
We elaborate on this in Sec.~\ref{Sec:Discussion}.

\subsection{Comparison of HEOM results and TLS predictions}
\label{SSec:HEOM-vs-TLS}

In the following, we compare and contrast TLS predictions with our HEOM results.
Figure~\ref{Fig:mu_dc_vs_T_w0_0.044_300924}(a) shows that the agreement of the TLS mobilities ($\alpha_d=2.2$) with the corresponding numerically exact results improves with increasing the temperature or the interaction.
In the weak-interaction regime $\lambda=0.05$, the TLS yields accurate mobilities at extremely high temperatures $T/\omega_0\gtrsim 20$, at which the Boltzmann picture of weak occasional carrier--phonon scatterings breaks down due to the large number of incoherent phonons.
As the temperature is lowered, the quality of the TLS predictions deteriorates, and the Boltzmann picture is restored. 
For moderate interactions $\lambda=0.336$ and 0.5, the Boltzmann theory largely overestimates the numerically exact $\mu_\mathrm{dc}(T)$, while the TLS reproduces it very well throughout the temperature interval examined.
Some differences between the TLS and HEOM mobilities appear near the lower end of the temperature range examined (at $T/J=0.1$ and 0.175).
These differences are within or somewhat above the ten-percent relative uncertainty accompanying HEOM results.
In this weak-disorder regime, the predictive power of the TLS in mobility computations could be improved by modifying it so that the modification interpolates between the Boltzmann transport theory and the standard TLS [Eq.~\eqref{Eq:TLS_ansatz}]~\cite{PhysRevResearch.2.013001}.
The rationale behind the success of this TLS modification was that QMC results were by $\sim 15\%$ lower than Boltzmann results and by $\sim 15\%$ higher than TLS predictions, see Fig.~1(a) of Ref.~\onlinecite{PhysRevResearch.2.013001}. 
However, Fig.~\ref{Fig:mu_dc_vs_T_w0_0.044_300924}(a) shows that the differences between HEOM and Boltzmann results are much larger ($\sim 50\%$) than the differences between HEOM and TLS results, while we have already established a good agreement between HEOM and QMC results.
This is discussed in more detail in Sec.~SIII of the Supplemental Material~\cite{comment261224}.

Figure~\ref{Fig:diffusion_exponents_230924} reveals qualitative differences between the TLS and HEOM dynamics of $\mathcal{D}$ and $\alpha$ for $\lambda=0.336$ and $T/J=0.175$ and 0.1.
Just as the imaginary-axis QMC, the TLS does not capture the transient nature of the subdiffusive carrier dynamics at $T/J=0.175$.
While the TLS and HEOM results for $\mathcal{D}(t)$ and $\alpha(t)$ agree very well for $t\lesssim t_\mathrm{min}$, the TLS predicts that the diffusive transport is approached from the subdiffusive side, in contrast to our HEOM results.
Although the TLS correctly predicts that the diffusive transport sets in from the superdiffusive side at $T/J=0.1$, it does not capture the weakly pronounced transient slowdown of the carrier (a dip in $\mathcal{D}$ and plateau in $\alpha$) occurring around $Jt\sim 50$.
At higher temperatures ($T/J=0.5$ and 1), the TLS results almost fully coincide with the numerically exact ones. 

\begin{figure}
    \centering
    \includegraphics[width=\columnwidth]{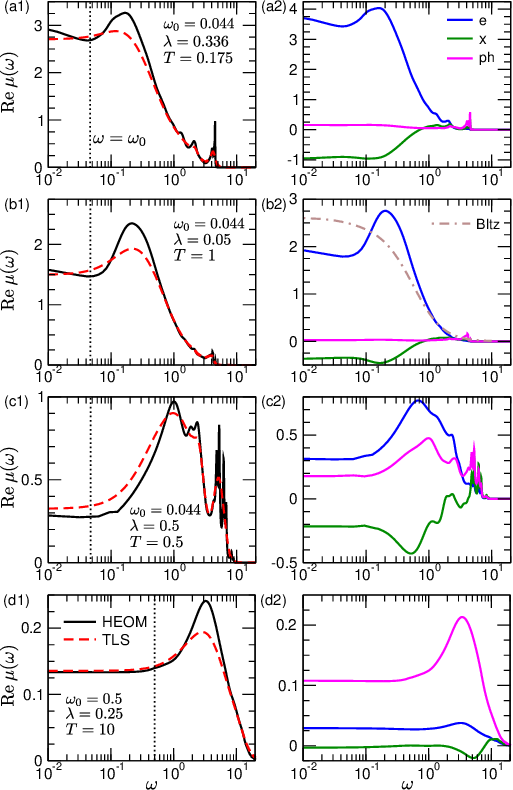}
    \caption{(a1)--(d1): Dynamical-mobility profiles computed using HEOM (solid lines), TLS (dashed lines), and Boltzmann theory [the dash-dotted line in (b2)] in different parameter regimes. (a2)--(d2): Purely electronic (label "e"), cross (label "x"), and phonon-assisted (label "ph") contributions to the HEOM dynamical-mobility profiles in (a1)--(d1).
    The dotted lines in (a1)--(d1) display $\omega=\omega_0$.
    We set $J=1$ in all panels.}
    \label{Fig:optical_conductivities_141024}
\end{figure}

Figure~\ref{Fig:optical_conductivities_141024} analyzes typical dynamical-mobility profiles and different contributions to them.
For $\lambda=0.336$ and $T/J=0.175$, the TLS closely follows the HEOM dynamical-mobility profile for $\omega/J\gtrsim 0.3$, reproducing the high-frequency feature around $\omega/J\approx 4$ that stems from the phonon-assisted contribution, see Figs.~\ref{Fig:optical_conductivities_141024}(a1) and~\ref{Fig:optical_conductivities_141024}(a2).
The TLS captures reasonably well the position of the HEOM-profile DDP centered around $\omega_\mathrm{DDP}/J\approx 0.2$, which mainly originates from the purely electronic contribution, see Fig.~\ref{Fig:optical_conductivities_141024}(a2).
However, the TLS does not fully capture the peak shape and intensity, misses the dip around $\omega_\mathrm{dip}\approx\omega_0$, and predicts that $\omega=0$ is a local minimum (instead of a local maximum) of $\mathrm{Re}\:\mu(\omega)$.
The situation is similar upon reducing $\lambda$ and increasing $T$ by approximately the same factor, so that the dynamical-disorder strength $\sigma^2=2\lambda JT$ remains approximately constant, see Fig.~\ref{Fig:optical_conductivities_141024}(a2).
Then, despite weak interaction, the Boltzmann picture of occasional carrier--phonon scatterings, whose formal expression is the computation of $\langle j_\mathrm{e}(t)j_\mathrm{e}(0)\rangle$ in the single-particle (bubble) approximation, cannot recover the numerically exact purely electronic contribution to $\mathrm{Re}\:\mu(\omega)$, see Fig.~\ref{Fig:optical_conductivities_141024}(b2).
Even though the cross contribution is appreciable, Figs.~\ref{Fig:optical_conductivities_141024}(a) and~\ref{Fig:optical_conductivities_141024}(b) show that it changes the purely electronic contribution only quantitatively, and not qualitatively.
This observation is further corroborated by the fact that $\mathrm{Re}\:\mu_\mathrm{e}(\omega)$ in Figs.~\ref{Fig:optical_conductivities_141024}(a2) and~\ref{Fig:optical_conductivities_141024}(b2) satisfies the "partial optical sum rule"
\begin{equation}
    \int_0^{+\infty}d\omega\:\mathrm{Re}\:\mu_\mathrm{e}(\omega)=-\frac{\pi}{2}\langle H_\mathrm{e}\rangle,
\end{equation}
which cannot be rigorously derived, with relative accuracy of the order of $10^{-3}$.
At the same time, the (full) optical sum rule $\int_0^{+\infty}d\omega\:\mathrm{Re}\:\mu(\omega)=-\frac{\pi}{2}\langle H_\mathrm{e}+H_\mathrm{e-ph}\rangle$ is satisfied with the relative accuracy of the order of $10^{-4}$, see Appendix~\ref{App:finite-N-and-D}.
Therefore, up to the non-negligible cross contribution, the overall physical situation in Figs.~\ref{Fig:optical_conductivities_141024}(a) and~\ref{Fig:optical_conductivities_141024}(b) is analogous to that we have recently analyzed in the Holstein model~\cite{JChemPhys.159.094113,PhysRevB.109.214312}, in which the current operator is purely electronic.
There, we have concluded that the finite-frequency peak in the carriers' optical response can be reproduced only by theories that compute $\langle j_\mathrm{e}(t)j_\mathrm{e}(0)\rangle$ without invoking the single-particle approximation, i.e., take vertex corrections into account.

The DDPs in Figs.~\ref{Fig:optical_conductivities_141024}(a1) and~\ref{Fig:optical_conductivities_141024}(b1) reflect carrier dynamics happening well before a single phonon period, when phonon dynamics can be safely ignored, as assumed within the TLS.
However, the TLS ansatz [Eq.~\eqref{Eq:TLS_ansatz}] is too simple to fully take into account the nontrivial charge--phonon dynamics on longer timescales (approaching one phonon period), when phonon motions cannot be neglected.
Computing such dynamics is generally an arduous task, which fortunately does not have to be performed at sufficiently high temperatures or for sufficiently strong interactions, when the diffusive transport is expected to set in before the first phonon period. 
Then, the TLS is expected to reproduce the numerically exact dynamical mobility very well, which is indeed confirmed in Fig.~\ref{Fig:optical_conductivities_141024}(c1).
Interestingly, even though the phonon-assisted contribution to $\mu_\mathrm{dc}$ is non-negligible, it is almost exactly canceled by the corresponding cross contribution, see Fig.~\ref{Fig:optical_conductivities_141024}(c2), so that $\mu_\mathrm{dc}$ is effectively determined only by the purely electronic contribution.
However, the phonon-assisted contribution to $\mu_\mathrm{dc}$ generally becomes more important as $\lambda$ or $T$ are increased, see Fig.~\ref{Fig:mu_dc_vs_T_w0_0.044_300924}(b) and the companion paper~\cite{part1}.
Ultimately, we expect the TLS to excel in the regime of phonon-assisted transport.
While this regime is difficult to reach for $\omega_0/J=0.044$, see Fig.~\ref{Fig:mu_dc_vs_T_w0_0.044_300924}(b), the results in our companion paper~\cite{part1} suggest that it is more easily reached for faster phonons. 
Figure~\ref{Fig:optical_conductivities_141024}(d1) shows that the TLS works well even when the timescales of free-phonon and free-carrier dynamics are comparable to one another ($\omega_0/J=0.5$), as long as the phonon-assisted contribution dominates the transport, see Fig.~\ref{Fig:optical_conductivities_141024}(d1).
Therefore, the timescale separation between carriers and phonons may not be so essential a criterion for the applicability of the TLS.
Figure~\ref{Fig:optical_conductivities_141024}(d1) suggests that it may be more important that the diffusive transport be reached (well) before the first phonon period, so that phonons are to a very good approximation frozen over the time window in which $C_{jj}(t)$ is appreciable.
In Sec.~SIV of the Supplemental Material~\cite{comment261224}, we complement Fig.~\ref{Fig:optical_conductivities_141024}(d) by an extensive comparison of HEOM results and TLS predictions for $\omega_0/J=0.5$.

Finally, we note that the results of the most recent quantum--classical approaches to the coupled carrier--phonon dynamics~\cite{runeson2024} bear qualitative similarity to our fully quantum results for 
$\mathrm{Re}\:\mu(\omega)$ in Fig.~\ref{Fig:optical_conductivities_141024}(a1).
In Sec.~SV of the Supplemental Material~\cite{comment261224}, we focus on the parameter regime $\omega_0/J=0.044,\lambda=0.5,T/J=0.175$, and establish a good agreement between the quantum--classical~\cite{runeson2024} and our fully quantum carrier mobilities, while we find some differences in the positions of the low-frequency features of the two dynamical-mobility profiles.
As this parameter regime pushes our HEOM-based methodology to its limits of applicability, we cannot definitely attribute the above-mentioned differences to the quantum--classical approximation. 

\section{Discussion}
\label{Sec:Discussion}

Both the TLS~\cite{PhysRevB.83.081202,AdvFunctMater.26.2292} and the treatment of phonons as a viscous medium~\cite{PhysRevLett.114.086601} introduce exponentially decaying terms in the dynamics, the former through Eq.~\eqref{Eq:TLS_ansatz}, and the latter through the friction force proportional to the velocity of the harmonically bound carrier (see the Supplemental Material to Ref.~\onlinecite{PhysRevLett.114.086601}).
Although the exponential-decay timescale $\tau_d\propto\omega_0^{-1}$ in Eq.~\eqref{Eq:TLS_ansatz} is established on the basis of physical arguments~\cite{PhysRevB.83.081202,AdvFunctMater.26.2292}, we note that similar ans\"{a}tze are often used to effectively take into account the influence of other scattering mechanisms not included in the model [Eqs.~\eqref{Eq:H_tot}--\eqref{Eq:H_ph}] considered.

For example, the polaron transformation-based approaches~\cite{JChemPhys.83.1854,PhysRevB.69.075212,JPhysChemB.115.5312} assume that
\begin{equation}
\label{Eq:PT_ansatz}
    C_{jj}(t)=C_{jj}^\mathrm{PT}(t)\:e^{-(\Gamma t)^2},
\end{equation}
where $C_{jj}^\mathrm{PT}(t)$ is the current--current correlation function evaluated in the polaron frame, whereas the phenomenological parameter $\Gamma$ characterizes the effective line broadening due to scattering mechanisms that are slow with respect to those determining the dynamics of $C_{jj}^\mathrm{PT}$.
This so-called inhomogeneous broadening~\cite{PhysRevB.69.075212,Mukamel-book,May-Kuhn-book,Valkunas-Abramavicius-Mancal-book}, which is characterized by the Gaussian damping in the time domain, is usually attributed to the presence of static disorder.
The exponential damping in the time domain [Eq.~\eqref{Eq:TLS_ansatz}] is, on the other hand, characteristic of the homogeneous broadening~\cite{Mukamel-book,May-Kuhn-book,Valkunas-Abramavicius-Mancal-book}, which is due to scattering mechanisms that are fast with respect to those taken into account in $C_{jj}^\mathrm{dis}(t)$.
The homogeneous broadening is often associated with the interaction with phonons~\cite{Mukamel-book,May-Kuhn-book,Valkunas-Abramavicius-Mancal-book}, which is indeed fast with respect to the scattering on static disorder that determines $C_{jj}^\mathrm{dis}(t)$.
Therefore, the TLS ansatz [Eq.~\eqref{Eq:TLS_ansatz}] may be thought of as a way to include scattering on additional phonon modes not explicitly considered in the model. 

From that viewpoint, the qualitative differences between the numerically exact and TLS results may be due to the fact Eq.~\eqref{Eq:TLS_ansatz} effectively takes into account interactions of the carrier with additional phonon modes not considered in the model [Eqs.~\eqref{Eq:H_tot}--\eqref{Eq:H_ph}].
Such a possibility is supported by the most recent quantum--classical simulations~\cite{runeson2024}.
There, the authors conclude that considering a continuous phonon spectrum centered around $\omega_0$ instead of the delta-like spectrum used here [see Eq.~\eqref{Eq:H_ph}] tends to diminish the long-time growth of $\mathcal{D}(t)$ and consequently the low-frequency features of the dynamical mobility.
While the finite-frequency peak in $\mathrm{Re}\:\mu(\omega)$ remains largely unaffected upon replacing the discrete phonon spectrum by the continuous one, the dip around $\omega_0$ and the zero-frequency peak become less pronounced with increasing the width of the phonon spectrum, see Fig.~4(b) of Ref.~\onlinecite{runeson2024}.
In other words, the dynamical-mobility profile bears stronger qualitative resemblance to TLS predictions.
In a similar vein, the long-time growth of $\mathcal{D}(t)$ is suppressed with increasing the width of the phonon spectrum, see Fig.~4(a) of Ref.~\onlinecite{runeson2024}.
Keeping in mind the above-established qualitative agreement between HEOM and quantum--classical approaches, one can expect that the rich structure of the numerically exact dynamical-mobility profiles in Figs.~\ref{Fig:optical_conductivities_141024}(a1) and~\ref{Fig:optical_conductivities_141024}(b1) would be less pronounced when considering a more realistic spectral density of the carrier--phonon interaction, which conveniently combines information on the interaction constants and phonon density of states~\cite{Mukamel-book}.

The authors of Ref.~\onlinecite{PhysRevLett.102.116602} showed that the realistic spectral density in partially ordered organic semiconductors can be approximated by a superposition of a small number of Langevin (or underdamped Brownian~\cite{Mukamel-book}) oscillators.
In numerically exact approaches based on the theory of open quantum systems~\cite{JChemPhys.132.081101,JChemPhys.151.044105,JChemPhys.161.084118}, such spectral densities are treated in the same manner as the more widely used Drude--Lorentz (or overdamped Brownian oscillator~\cite{Mukamel-book}) spectral density~\cite{JChemPhys.140.134106}.
Nevertheless, as noted in Sec.~\ref{Sec:Introduction}, such approaches have not been applied in numerically exact mobility computations based on $C_{jj}(t)$.
Concerning the HEOM-based approaches~\cite{JChemPhys.132.081101} dealing with continuous spectral densities, possible reasons behind that state of affairs are the issues with the phonon-assisted current, and the well-known incompatibility of the hierarchies in the imaginary-time and real-time domains~\cite{JChemPhys.141.044114}.
Our study could motivate an extension of the approach used here to the case of continuous spectral densities, which would rigorously prove or disprove the claims of this section.
In the meantime, the most recent numerically exact~\cite{JChemTheoryComput.20.7052} and approximate~\cite{ChemSci.15.16715} quantum-dynamics computations of transport properties of the Holstein model with overdamped and/or underdamped Brownian oscillator spectral density show that $\mathcal{D}(t)$ and $\mathrm{Re}\:\mu(\omega)$ bear stronger qualitative resemblance to the TLS predictions~\cite{PhysRevLett.132.266502} than to the numerically exact results~\cite{JChemPhys.159.094113,PhysRevB.72.104304,PhysRevB.106.155129,mitric2024dynamicalquantumtypicalitysimple,mitric2024precursorsandersonlocalizationholstein} that assume discrete undamped phonons.    

The model considered here also does not take into account the extrinsic static disorder.
Quantum--classical simulations in Ref.~\onlinecite{runeson2024} suggest that such static disorder in on-site energies diminishes the long-time growth of the diffusion constant, rendering it overall qualitatively similar to the TLS predictions, see Fig.~5 of Ref.~\onlinecite{runeson2024}.
This once again suggests that the qualitative differences between the numerically exact and TLS results could be ascribed to the too simple model we consider.
In more realistic models considering the interactions with additional phonon modes and static disorder, one expects that the true dynamical-mobility profile is qualitatively similar to the TLS prediction.

\section{Conclusion}
\label{Sec:Conclusion}
Our study provides the long-awaited quantum-dynamical insights into transport properties of the one-dimensional Peierls model with a single undamped vibration per lattice site in the adiabatic regime.
For the parameters representative of the room-temperature transport in crystalline rubrene, we establish that the crossover from super- to subdiffusive carrier dynamics is of transient nature, so that the long-time diffusive transport is eventually approached from the superdiffusive side on timescales of the order of one phonon period.
Our findings stand in qualitative contrast to those of the best available numerically exact (imaginary-axis QMC) and the most widely used approximate (TLS) methods, both of which conclude that the diffusive transport sets in from the subdiffusive side.
On the other hand, our transport dynamics strongly supports the results of the most recent quantum--classical simulations.

For the most widely studied combinations of model parameters, the TLS can reproduce HEOM mobilities very well once the free-parameter $\alpha_d\sim 1$ is appropriately tuned.
Our results thus suggest that any approach that reasonably captures carrier dynamics on short to intermediate timescales can be expected to yield reasonable predictions for the carrier mobility, even though it may poor at treating the long-time coupled carrier--phonon dynamics.
At higher temperatures or for stronger interactions, the exact form of these long-time dynamics is immaterial because the carrier diffusion is established well before a single phonon period, so that the frozen-phonon approximation is reasonable whenever the current--current correlation function is appreciable.
In such situations, the TLS ansatz in Eq.~\eqref{Eq:TLS_ansatz} is sufficiently good to describe this predominantly phonon-assisted transport, even when the timescales of carrier and phonon dynamics are comparable.  
At realistic temperatures and interactions, we argue that Eq.~\eqref{Eq:TLS_ansatz} can be considered to effectively take into account other scattering mechanisms not included in the present model.
Apart from providing numerically exact results for carrier dynamics, this piece of research can be regarded as a formal justification of the already well-established practical applicability of the TLS to realistic systems, for which the model embodied in Eqs.~\eqref{Eq:H_tot}--\eqref{Eq:H_ph} is too simplistic. 

\begin{acknowledgments}
This research was supported by the Science Fund of the Republic of Serbia, Grant No. 5468, Polaron Mobility in Model Systems and Real Materials--PolMoReMa.
The author acknowledges funding provided by the Institute of Physics Belgrade through a grant from the Ministry of Science, Technological Development, and Innovation of the Republic of Serbia.
Numerical computations were performed on the PARADOX-IV supercomputing facility at the Scientific Computing Laboratory, National Center of Excellence for the Study of Complex Systems, Institute of Physics Belgrade.
The author thanks Nenad Vukmirovi\'c for many useful and stimulating discussions.
\end{acknowledgments}

\section*{Data Availability Statement}
The data that support the findings of this article are openly available~\cite{jankovic_2025_14637019}.

\newpage
\begin{widetext}
\appendix
\section{Finite-size and finite-depth effects in HEOM computations for $\omega_0/J=0.044,\lambda=0.336,$ and $T/J=0.175$}
\label{App:finite-N-and-D}
Here, we consider in greater detail the effects due to finite values of $N$ and $D$ in HEOM computations in the parameter regime representative of the room-temperature carrier transport along the direction of maximal conductivity in rubrene.
\begin{figure}[htbp!]
    \centering
    \includegraphics[width=0.75\textwidth]{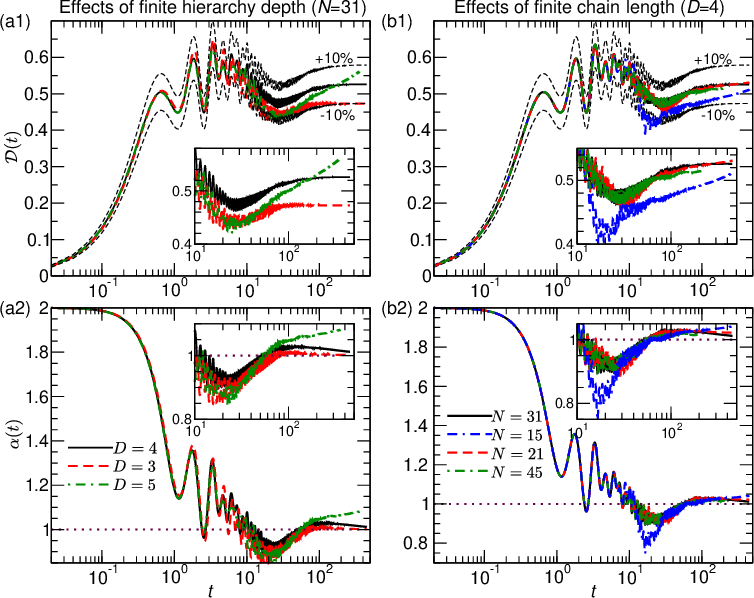}
    \caption{Effects of finite maximum hierarchy depth $D$ [(a1) and (a2)] and finite chain length $N$ [(b1) and (b2)] on the dynamics of the diffusion constant [(a1) and (b1)] and diffusion exponent [(a2) and (b2)].
    In (a1) and (a2), the chain length is set to $N=31$.
    In (b1) and (b2), the maximum hierarchy depth is set to $D=4$.
    The insets zoom in the dynamics of $\mathcal{D}$ and $\alpha$ on intermediate to long time scales.
    The dashed lines in (a1) and (b1) represent the quantities $1.1\times\mathcal{D}_{(31,4)}(t)$ (label "$+10\%$") and $0.9\times\mathcal{D}_{(31,4)}(t)$ (label "$-10\%$"), where $\mathcal{D}_{(31,4)}(t)$ is the diffusion constant computed for $N=31$ and $D=4$.
    The model parameters assume the following values: $J=1,\omega_0=0.044,\lambda=0.336,$ and $T=0.175$.}
    \label{Fig:finite_N_D}
\end{figure}
We examine the dynamics of the diffusion constant [Figs.~\ref{Fig:finite_N_D}(a1) and~\ref{Fig:finite_N_D}(b1)] and the diffusion exponent [Figs.~\ref{Fig:finite_N_D}(a2) and~\ref{Fig:finite_N_D}(b2)] for different maximum hierarchy depths $D$ [fixing $N=31$, see Figs.~\ref{Fig:finite_N_D}(a1) and~\ref{Fig:finite_N_D}(a2)] and for different chain lengths $N$ [fixing $D=4$, see Figs.~\ref{Fig:finite_N_D}(b1) and~\ref{Fig:finite_N_D}(b2)].

Fixing $N=31$, we find that the dc mobility can be reliably extracted from $\mathcal{D}(t)$ computed for $D=3$ and 4, see Fig.~\ref{Fig:finite_N_D}(a1).
On the other hand, $\mathcal{D}(t)$ for $D=5$ does not exhibit the long-time saturation, which prevents us from reliably estimating $\mu_\mathrm{dc}$.
On the level of the diffusion exponent, the results for $D=3$ and $D=4$ exhibit the expected long-time approach towards unity, while the long-time behavior of the result for $D=5$ is qualitatively wrong, see Fig.~\ref{Fig:finite_N_D}(a2).
Such a behavior is a consequence of the ineffectiveness of our HEOM closing strategy when the maximum hierarchy depth is sufficiently large, see Sec.~V.C of the companion paper~\cite{part1}. 
Nevertheless, the dynamics of $\mathcal{D}$ and $\alpha$ on short to intermediate timescales are qualitatively (and to a large extent quantitatively) similar for all three values of $D$ studied.
In particular, upon reaching their minima on intermediate timescales, both $\mathcal{D}(t)$ and $\alpha(t)$ increase on longer timescales.
This increase is more pronounced for $D=4$ than for $D=3$, so that the approach towards the long-time diffusive transport from the superdiffusive side is not so convincing for $D=3$.
Analyzing the relative error
\begin{equation}
\label{Eq:delta_OSR}
    \delta_\mathrm{OSR}=\frac{\left|\int_0^{+\infty}d\omega\:\mathrm{Re}\:\mu(\omega)+\frac{\pi}{2}\langle H_\mathrm{e}+H_\mathrm{e-ph}\rangle\right|}{\frac{\pi}{2}|\langle H_\mathrm{e}+H_\mathrm{e-ph}\rangle|}   
\end{equation}
with which the optical sum rule is satisfied, see Table~\ref{Tab:delta-OSR_vs_D_N_31}, we conclude that the result for $D=4$ is more reliable than that for $D=3$.
However, we point out that, close to the adiabatic limit, the smallness of $\delta_\mathrm{OSR}$ does not guarantee the absolute reliability of the corresponding result.
Namely, as the adiabaticity ratio $\omega_0/J$ is decreases, it becomes increasingly difficult to converge the results for the thermodynamic expectation value $\langle H_\mathrm{e}+H_\mathrm{e-ph}\rangle$ that determines $\delta_\mathrm{OSR}$.
This is clearly seen in Table~\ref{Tab:delta-OSR_vs_D_N_31}, which shows that $\delta_\mathrm{OSR}$ for $D=3$ and 4 is better than the number of significant figures in $\langle H_\mathrm{e}+H_\mathrm{e-ph}\rangle$.
The error with which the symmetry $\langle j_\mathrm{e}(t)j_\mathrm{e-ph}(0)\rangle=\langle j_\mathrm{e-ph}(t)j_\mathrm{e}(0)\rangle$ (see Sec.~IV.C of the companion paper~\cite{part1}) is obeyed is of the order of $10^{-2}$ for all three values of $D$.
The HEOM mobilities for $D=3$ and $D=4$ differ by around $10\%$. 
\begin{table}[htbp!]
    \centering
    \begin{tabular}{c|c|c|c|c}
        $D$ & $t_\mathrm{max}$ & $\delta_\mathrm{OSR}$ & $-\langle H_\mathrm{e}+H_\mathrm{e-ph}\rangle$ & $\max\left|\left\langle j_\mathrm{e}(t)j_\mathrm{e-ph}(0)-j_\mathrm{e-ph}(t)j_\mathrm{e}(0)\right\rangle\right|$\\\hline\hline
        3 & 450 & $5.1\times 10^{-3}$ & $2.0630327684$ & $8.0\times 10^{-3}$ \\
        \textbf{4} & \textbf{450} & $\mathbf{1.0\times 10^{-4}}$ & $2.0674819496$ & $9.0\times 10^{-3}$ \\
        5 & 350 & $2.3\times 10^{-4}$ & $2.0687529991$ & $6.4\times 10^{-3}$
    \end{tabular}
    \caption{Dependence of $\delta_\mathrm{OSR}$ [Eq.~\eqref{Eq:delta_OSR}], $\langle H_\mathrm{e}+H_\mathrm{e-ph}\rangle$, and $\max\left|\left\langle j_\mathrm{e}(t)j_\mathrm{e-ph}(0)-j_\mathrm{e-ph}(t)j_\mathrm{e}(0)\right\rangle\right|$ on $D$ for $N=31$. The model parameters are set to $J=1,\omega_0=0.044,\lambda=0.336,$ and $T=0.175$.}
    \label{Tab:delta-OSR_vs_D_N_31}
    \centering
    \begin{tabular}{c|c|c}
        $N$ & $t_\mathrm{max}$ & $\delta_\mathrm{OSR}$\\\hline\hline
        15 & 450 & $1.5\times 10^{-3}$ \\
        21 & 450 & $1.9\times 10^{-4}$ \\
        \textbf{31} & \textbf{450} & $\mathbf{1.0\times 10^{-4}}$ \\
        45 & 205 & $9.5\times 10^{-5}$
    \end{tabular}
    \caption{Dependence of $\delta_\mathrm{OSR}$ [Eq.~\eqref{Eq:delta_OSR}] on $N$ for $D=4$. The model parameters are set to $J=1,\omega_0=0.044,\lambda=0.336,$ and $T=0.175$.}
    \label{Tab:delta-OSR_vs_N_D_4}
\end{table}

Fixing $D=4$, we find that the result for $N=15$ displays finite-size effects, while the results for $N=21,31,$ and 45 are qualitatively and to a large extent quantitatively similar, see Figs.~\ref{Fig:finite_N_D}(b1) and~\ref{Fig:finite_N_D}(b2).
The lower quality of the result for $N=15$ with respect to the results on longer chains is further corroborated by Table~\ref{Tab:delta-OSR_vs_N_D_4}, which shows that $\delta_\mathrm{OSR}$ is of the same order of magnitude for $N=21,31,$ and 45.
Due to the large computational cost of the HEOM calculations for $N=45$, we stopped them as soon as we certified that the corresponding dynamics of $\mathcal{D}$ and $\alpha$ (and consequently the dc mobility) are very close to those on shorter chains ($N=31$ and 21).
We note that a fully reliable identification of the dip of the dynamical-mobility profile at $\omega\approx\omega_0$ necessitates simulation times that are longer than a single phonon period $2\pi/\omega_0=143$.
Our reference HEOM computations ($N=31,D=4$) are performed up to $Jt=450$, which is somewhat longer than 3 phonon periods, once again suggesting that the existence of the dip in $\mathrm{Re}\:\mu(\omega)$ at $\omega\approx\omega_0$ can be considered as reliably established.

\section{Effects of closing scheme in HEOM computations for $\omega_0/J=0.044,\lambda=0.336,$ and $T/J=0.175$}
\label{App:closing-scheme}
In Appendix~\ref{App:finite-N-and-D}, we establish that $N=31$ and $D=4$ yield the best available HEOM results for $\omega_0/J=0.044,\lambda=0.336,$ and $T/J=0.175$.
Here, we discuss the influence of the HEOM closing scheme, which represents the main approximation of our HEOM-based approach, on the results for $N=31$ and $D=4$.
\begin{figure}[htbp!]
    \centering
    \includegraphics[width=0.75\textwidth]{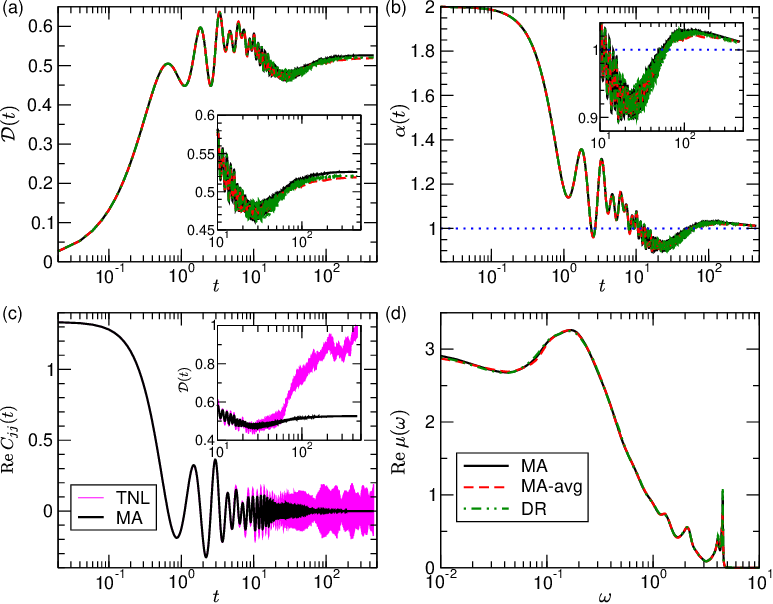}
    \caption{
    (a)--(c): Dynamics of (a) the diffusion constant, (b) the diffusion exponent, and (c) the real part of the current autocorrelation function for different hierarchy closing schemes.
    (d): Frequency-dependent mobility for different hierarchy closing schemes.
    In (a), (b), and (d), we compare the results obtained using the Markovian-adiabatic scheme (label "MA", solid lines), the Markovian-adiabatic scheme with momentum-averaged quasiparticle scattering rates (label "MA-avg", dashed lines), and the derivative-resum scheme (label "DR", dash-dotted lines).
    In (c), we compare the MA (thick line) to the time-nonlocal (label "TNL", thin line) scheme, which sets closing terms to zero.
    The insets in (a), (b), and (c) display the dynamics of $\mathcal{D}$, $\alpha$, and $\mathcal{D}$, respectively, on intermediate-to-long time scales.
    }
    \label{Fig:complement_pII_fig_3a}
\end{figure}

Figures~\ref{Fig:complement_pII_fig_3a}(a) and~\ref{Fig:complement_pII_fig_3a}(b) compare the dynamics of the diffusion constant and exponent for different closing schemes discussed in Sec.~IV.C of the companion paper~\cite{part1}.
Apart from the Markovian-adiabatic scheme, we consider the Markovian-adiabatic scheme with momentum-averaged quasiparticle scattering rates, and the derivative-resum scheme.
The overall dynamics of $\mathcal{D}$ and $\alpha$ is essentially independent of the particular closing scheme.
Comparing the insets of Figs.~\ref{Fig:complement_pII_fig_3a}(a) and~\ref{Fig:complement_pII_fig_3a}(b) to the insets of Figs.~\ref{Fig:finite_N_D}(a1)--\ref{Fig:finite_N_D}(b2), we conclude that the differences between the results relying on different closing schemes for fixed $N$ and $D$ are smaller than the differences between the results for different values of $N$ and $D$ and fixed closing scheme.
The frequency-dependent mobility is also very weakly dependent on the closing scheme, see Fig.~\ref{Fig:complement_pII_fig_3a}(d).

Figure~\ref{Fig:complement_pII_fig_3a}(c) and its inset, which compare $\mathrm{Re}\:C_{jj}(t)$ and $\mathcal{D}(t)$ computed using Markovian-adiabatic closing terms and zero closing terms (the so-called time-nonlocal scheme), emphasize that the hierarchy closing is vital to computing long-time transport dynamics and carrier mobility.
The closing-induced stabilization of carrier dynamics sets in for $t\gtrsim 1/\omega_0$, and prevents long-time oscillations of $\mathrm{Re}\:C_{jj}(t)$ around zero [the main panel of Fig.~\ref{Fig:complement_pII_fig_3a}(c)] that are responsible for the long-time increase of $\mathcal{D}(t)$ [the inset of Fig.~\ref{Fig:complement_pII_fig_3a}(c)].
The dynamics without closing remains reasonable up to around one half of the phonon period (for $1/\omega_0\lesssim t\lesssim\pi/\omega_0$), when it qualitatively resembles the dynamics relying on the MA closing.
In particular, both curves in the inset of Fig.~\ref{Fig:complement_pII_fig_3a}(c) show that the diffusion constant increases after reaching its local minimum at $t_\mathrm{min}\approx 1/\omega_0$.
Therefore, the increase in $\mathcal{D}(t)$ for $t\gtrsim t_\mathrm{min}$ is not an artifact of the hierarchy closing.
The closing-induced stabilization of the long-time dynamics of $\mathcal{D}$ and $\alpha$ is very weakly dependent on the underlying approximations, and the corresponding carrier mobility agrees very well with the numerically exact results available in the literature.
\end{widetext}
\bibliography{aipsamp}

\end{document}


\title{Supplemental Material for\\
Charge transport limited by nonlocal electron--phonon interaction. II. Numerically exact quantum dynamics in the slow-phonon regime}

\author{Veljko Jankovi\'c}
 \email{veljko.jankovic@ipb.ac.rs}
 \affiliation{Institute of Physics Belgrade, University of Belgrade, Pregrevica 118, 11080 Belgrade, Serbia}
\maketitle

\section{Details of HEOM computations}
In this section, $N$ denotes the chain length, $D$ is the maximum hierarchy depth, $t_\mathrm{max}$ is the maximum (real) time up to which HEOM are propagated, while $\delta_\mathrm{OSR}$ [Eq.~(A1) of the main text] is the relative accuracy with which the optical sum rule is satisfied.
Our HEOM data are openly available in Ref.~\onlinecite{jankovic_2025_14637019}.
\begin{table}[htbp!]
    \centering
    \begin{tabular}{c|c|c|c|c|c|c}
    $\omega_0/J$ & $\lambda$ & $T/J$ & $N$ & $D$ & $t_\mathrm{max}$ & $\delta_\mathrm{OSR}$ \\\hline\hline
    0.044 & 0.05 & 0.1 & 201 & 2 & 2450 & $6.5\times 10^{-5}$\\\hline
    0.044 & 0.05 & 0.2 & 61 & 3 & 750 & $2.7\times 10^{-5}$\\\hline
    0.044 & 0.05 & 0.5 & 31 & 5 & 300 & $2.25\times 10^{-5}$\\\hline
    0.044 & 0.05 & 1 & 31 & 5 & 300 & $2.4\times 10^{-5}$\\\hline\hline
    0.044 & 0.336 & 0.1 & 71 & 3 & 621.64 & $4.0\times 10^{-3}$\\\hline
    0.044 & 0.336 & 0.175 & 31 & 4 & 450 & $1.0\times 10^{-4}$\\\hline
    0.044 & 0.336 & 0.5 & 21 & 6 & 100 & $1.3\times 10^{-4}$\\\hline
    0.044 & 0.336 & 1 & 10 & 8/9 & 50 & $8.5\times 10^{-5}$\\\hline\hline
    0.044 & 0.5 & 0.175 & 31 & 4 & 239.78 & $1.9\times 10^{-3}$\\\hline
    0.044 & 0.5 & 0.5 & 13 & 7/8 & 100 & $4.1\times 10^{-3}$\\\hline
    0.044 & 0.5 & 1 & 13 & 6/7 & 100 & $2.7\times 10^{-3}$\\\hline\hline
    \end{tabular}
    \caption{Details of the HEOM computations performed for $\omega_0/J=0.044$.}
    \label{Tab:details_w0_0.044}
\end{table}

\begin{table}[htbp!]
    \centering
    \begin{tabular}{c|c|c|c|c|c|c}
    $\omega_0/J$ & $\lambda$ & $T/J$ & $N$ & $D$ & $t_\mathrm{max}$ & $\delta_\mathrm{OSR}$ \\\hline\hline
    0.5 & 0.25 & 1 & 71 & 3 & 100 & $9.3\times 10^{-3}$\\\hline
    0.5 & 0.25 & 2 & 31 & 5 & 50 & $4.9\times 10^{-4}$\\\hline
    0.5 & 0.25 & 5 & 9 & 8/9 & 50 & $1.3\times 10^{-4}$\\\hline
    0.5 & 0.25 & 10 & 7 & 8/9 & 50 & $1.55\times 10^{-4}$\\\hline\hline
    0.5 & 0.5 & 1 & 31 & 5 & 30 & $8.45\times 10^{-3}$\\\hline
    0.5 & 0.5 & 2 & 13 & 7 & 50 & $2.0\times 10^{-4}$\\\hline
    0.5 & 0.5 & 5 & 7 & 11/12 & 50 & $1.4\times 10^{-4}$\\\hline
    0.5 & 0.5 & 10 & 7 & 11/12 & 50 & $1.6\times 10^{-4}$\\\hline\hline
    0.5 & 1 & 1 & 9 & 10 & 20 & $3.7\times 10^{-4}$\\\hline
    0.5 & 1 & 2 & 9 & 9/10 & 25 & $2.2\times 10^{-4}$\\\hline
    0.5 & 1 & 5 & 7 & 11/12 & 20 & $3.9\times 10^{-4}$\\\hline
    0.5 & 1 & 10 & 6 & 13/14 & 20 & $6.2\times 10^{-4}$\\\hline\hline
    \end{tabular}
    \caption{Details of the HEOM computations performed for $\omega_0/J=0.5$.}
    \label{Tab:details_w0_0.5}
\end{table}

\newpage

\section{Details of TLS computations}
\label{App:TLS_details}

We perform TLS computations on an $N$-site chain with open boundary conditions.
With this choice of boundary conditions, the matrix representing the Hamiltonian $H_\mathrm{e}+H_\mathrm{e-ph}$ is tridiagonal in the site basis, its elements on the upper and lower diagonal being $-J_{n,n+1}=-J+X_{n,n+1}$.
As in the main text, $X_{n,n+1}$ denotes a Gaussian random variable of zero mean and variance $\sigma^2=2\lambda JT$.
The tridiagonal property of the Hamiltonian matrix brings about computational savings when solving for its eigenvalues $\varepsilon_{d;x}$ and eigenvectors $|d;x\rangle=\sum_n\psi^{d;x}_n|n\rangle$ in each disorder realization (labeled by $d$).
The disorder-averaged current--current correlation function is then computed as
\begin{equation}
\label{Eq:C_jj_dis}
    C_{jj}^\mathrm{dis}(t)=\lim_{N_\mathrm{dis}\to+\infty}\frac{1}{N_\mathrm{dis}}\sum_{d=1}^{N_\mathrm{dis}}\sum_{x_2x_1}e^{i(\varepsilon_{d;x_2}-\varepsilon_{d;x_1})t}|\langle d;x_2|j|d;x_1\rangle|^2\frac{e^{-\beta\varepsilon_{d;x_2}}}{Z_d},
\end{equation}
where $Z_d=\sum_{x}e^{-\beta\varepsilon_{d;x}}$, while the current-operator matrix element is
\begin{equation}
    \langle d;x_2|j|d;x_1\rangle=-i\sum_{n=1}^{N-1}\left(-J+X_{n,n+1}^d\right)\left(\psi_n^{d;x_2*}\psi_{n+1}^{d;x_1}-\psi_{n+1}^{d;x_2*}\psi_n^{d;x_1}\right).
\end{equation}
Here, $X_{n,n+1}^d$ denotes the value of the random variable $X_{n,n+1}$ in $d$th disorder realization.
The averaging is performed over a sufficiently large number $N_\mathrm{dis}$ of disorder realizations.
We compute $C_{jj}^\mathrm{dis}(t)$ on a chain of length $N$ over the time window $[0,t_\mathrm{max}]$, where $N$ is sufficiently large and $t_\mathrm{max}$ is sufficiently long so that the absence of diffusion in the static-disorder setup is manifest, i.e., $\int_0^{t_\mathrm{max}}dt\:\mathrm{Re}\:C_{jj}^\mathrm{dis}(t)\approx 0$.
We summarize the values of model and numerical parameters of our TLS computations in Tables~\ref{Tab:TLS_details_0.044} (for $\omega_0/J=0.044$) and~\ref{Tab:TLS_details_0.5} (for $\omega_0/J=0.5$).
Our TLS data are openly available in Ref.~\onlinecite{jankovic_2025_14637019}.

If one is interested only in the TLS dc mobility, one can use Eq.~\eqref{Eq:C_jj_dis} to compute $\mu_\mathrm{dc}$ and obtain the well-known TLS expression~\cite{JPhysChemC.123.6989}
\begin{equation}
    \mu_\mathrm{dc}=\frac{1}{T}\int_0^{+\infty}dt\:e^{-t/\tau_d}C_{jj}^\mathrm{dis}(t)=\frac{1}{T}\frac{\overline{L^2}(\tau_d)}{2\tau_d},
\end{equation}
where the disorder-averaged squared transient localization length is defined as
\begin{equation}
    \overline{L^2}(\tau_d)=\lim_{N_\mathrm{dis}\to+\infty}\frac{1}{N_\mathrm{dis}}\sum_{d=1}^{N_\mathrm{dis}}\sum_{x_2x_1}|\langle d;x_2|j|d;x_1\rangle|^2\frac{2}{(\varepsilon_{d;x_2}-\varepsilon_{d;x_1})^2+\tau_d^{-2}}\frac{e^{-\beta\varepsilon_{d;x_2}}}{Z_d}.
\end{equation}

\begin{table}[htbp!]
    \centering
    \begin{tabular}{c|c|c|c|c|c}
        $\omega_0/J$ & $\lambda$ & $T/J$ & $N$ & $Jt_\mathrm{max}$ & $N_\mathrm{dis}$ \\
        \hline\hline
        0.044 & 0.05 & 0.2 & 768 & 1750 & 4096\\
        \hline
        0.044 & 0.05 & 0.5 & 512 & 1250 & 4096\\
        \hline
        0.044 & 0.05 & 1 & 384 & 1000 & 4096\\
        \hline\hline
        0.044 & 0.336 & 0.1 & 768 & 1250 & 6144\\
        \hline
        0.044 & 0.336 & 0.175 & 512 & 750 & 4096\\
        \hline
        0.044 & 0.336 & 0.5 & 512 & 750 & 4096\\
        \hline
        0.044 & 0.336 & 1 & 256 & 250 & 4096\\
        \hline\hline
        0.044 & 0.5 & 0.175 & 512 & 500 & 4096\\
        \hline
        0.044 & 0.5 & 0.5 & 512 & 250 & 4096\\
        \hline
        0.044 & 0.5 & 1 & 256 & 250 & 4096\\
        \hline\hline
    \end{tabular}
    \caption{Parameters used in TLS computations for the investigated values of $\lambda$ and $T/J$ for $\omega_0/J=0.044$.
    }
    \label{Tab:TLS_details_0.044}
\end{table}

\begin{table}[htbp!]
    \centering
    \begin{tabular}{c|c|c|c|c|c}
        $\omega_0/J$ & $\lambda$ & $T/J$ & $N$ & $Jt_\mathrm{max}$ & $N_\mathrm{dis}$ \\
        \hline\hline
        0.5 & 0.25 & 1 & 512 & 750 & 4096\\
        \hline
        0.5 & 0.25 & 2 & 512 & 500 & 4096\\
        \hline
        0.5 & 0.25 & 5 & 256 & 200 & 4096\\
        \hline
        0.5 & 0.25 & 10 & 256 & 200 & 4096\\
        \hline\hline
        0.5 & 0.5 & 1 & 512 & 400 & 4096\\
        \hline
        0.5 & 0.5 & 2 & 256 & 200 & 4096\\
        \hline
        0.5 & 0.5 & 5 & 256 & 150 & 4096\\
        \hline
        0.5 & 0.5 & 10 & 256 & 150 & 4096\\
        \hline\hline
        0.5 & 1 & 1 & 256 & 100 & 4096\\
        \hline
        0.5 & 1 & 2 & 256 & 100 & 4096\\
        \hline
        0.5 & 1 & 5 & 256 & 100 & 6144\\
        \hline
        0.5 & 1 & 10 & 256 & 100 & 6144\\
        \hline\hline
    \end{tabular}
    \caption{Parameters used in TLS computations for the investigated values of $\lambda$ and $T/J$ for $\omega_0/J=0.5$.
    }
    \label{Tab:TLS_details_0.5}
\end{table}

\newpage

\section{Boltzmann transport theory}
Within the Boltzmann transport theory, the carrier mobility is given as
\begin{equation}
\label{Eq:Boltzmann_most_general}
    \mu_\mathrm{dc}=\beta\sum_k v_k^2\tau_k\frac{e^{-\beta\varepsilon_k}}{Z},
\end{equation}
while the dynamical mobility can be computed as
\begin{equation}
\label{Eq:Boltzmann_mu_w}
    \mathrm{Re}\:\mu(\omega)=\beta\sum_k\frac{v_k^2\tau_k}{1+(\omega\tau_k)^2}\frac{e^{-\beta\varepsilon_k}}{Z}.
\end{equation}
In Eqs.~\eqref{Eq:Boltzmann_most_general} and~\eqref{Eq:Boltzmann_mu_w}, the wave number $k\in(-\pi,\pi]$ enumerates free-carrier states, $\varepsilon_k=-2J\cos k$ is the free-carrier dispersion, $Z=\sum_k e^{-\beta\varepsilon_k}$, $v_k=\frac{\partial\varepsilon_k}{\partial k}=2J\sin k$ is the carrier's band velocity, while $\tau_k$ is the relaxation time.
In the limit of low carrier density and in the momentum relaxation-time approximation, $\tau_k^\mathrm{MRTA}$ reads
\begin{equation}
\label{Eq:tau_k_MRTA}
\begin{split}
    \frac{1}{\tau_k^\mathrm{MRTA}}=\sum_q w_{k+q,k}\left(1-\cos\theta_{k+q,k}\right).
\end{split}
\end{equation}
In Eq.~\eqref{Eq:tau_k_MRTA}, $w_{k+q,k}$ is the transition rate from the free-electron state $|k\rangle$ to the free-electron state $|k+q\rangle$ in the second-order perturbation theory and reads
\begin{equation}
\label{Eq:w_k-plus-q_k_2nd}
    w_{k+q,k}=2\pi\frac{g^2}{N}|M(k,q)|^2\sum_{\pm}\left(n_\mathrm{ph}+\frac{1}{2}\pm\frac{1}{2}\right)\delta(\varepsilon_{k+q}-\varepsilon_k\pm\omega_0),
\end{equation}
where the carrier--phonon matrix element $M(k,q)$ is
\begin{equation}
    M(k,q)=-2i[\sin(k+q)-\sin k],
\end{equation}
while $n_\mathrm{ph}=\left(e^{\beta\omega_0}-1\right)^{-1}$ denotes the thermal occupation of free-phonon states labeled by the phonon wave number $q$.
The geometric factor in Eq.~\eqref{Eq:tau_k_MRTA} contains the cosine of the angle $\theta_{k+q,k}$ between the carrier velocities before and after its scattering on phonons:
\begin{equation}
    \cos\theta_{k+q,k}=\frac{v_{k+q}v_k}{|v_{k+q}||v_k|}.
\end{equation}
In the one-dimensional model considered here, $\theta_{k+q,k}$ can assume only the values of $0$ and $\pi$.

In the companion paper~\cite{part1}, we find that different widely used approximations to $\tau_k$ yield different results for $\mu_\mathrm{dc}$.
We do not expect these differences to be very pronounced in the slow-phonon limit $\omega_0/J=0.044$ considered in this study.
However, in view of the differences between the Boltzmann (in MRTA) and HEOM mobilities observed in Fig.~1(a) of the main text (for $\lambda=0.336$ and 0.5), we check this expectation by computing $\tau_k$ as the self-consistent solution to the following system of implicit equations:
\begin{equation}
\label{Eq:sc_tau_k_implicit}
    \frac{1}{\tau_k^\mathrm{SC}}=\sum_{q}w_{k+q,k}\left(1-\cos\theta_{k+q,k}\frac{|v_{k+q}|\tau_{k+q}^\mathrm{SC}}{|v_k|\tau_k^\mathrm{SC}}\right).
\end{equation}
We also consider the version of the MRTA used in Ref.~\onlinecite{PhysRevResearch.2.013001}, that we label MRTA1
\begin{equation}
\label{Eq:tau_k_MRTA1}
    \frac{1}{\tau_k^\mathrm{MRTA1}}=\sum_{q}w_{k+q,k}\left(1-\frac{v_{k+q}v_k}{|v_k|^2}\right).
\end{equation}

\begin{figure}[htbp!]
    \centering
    \includegraphics[width=0.75\columnwidth]{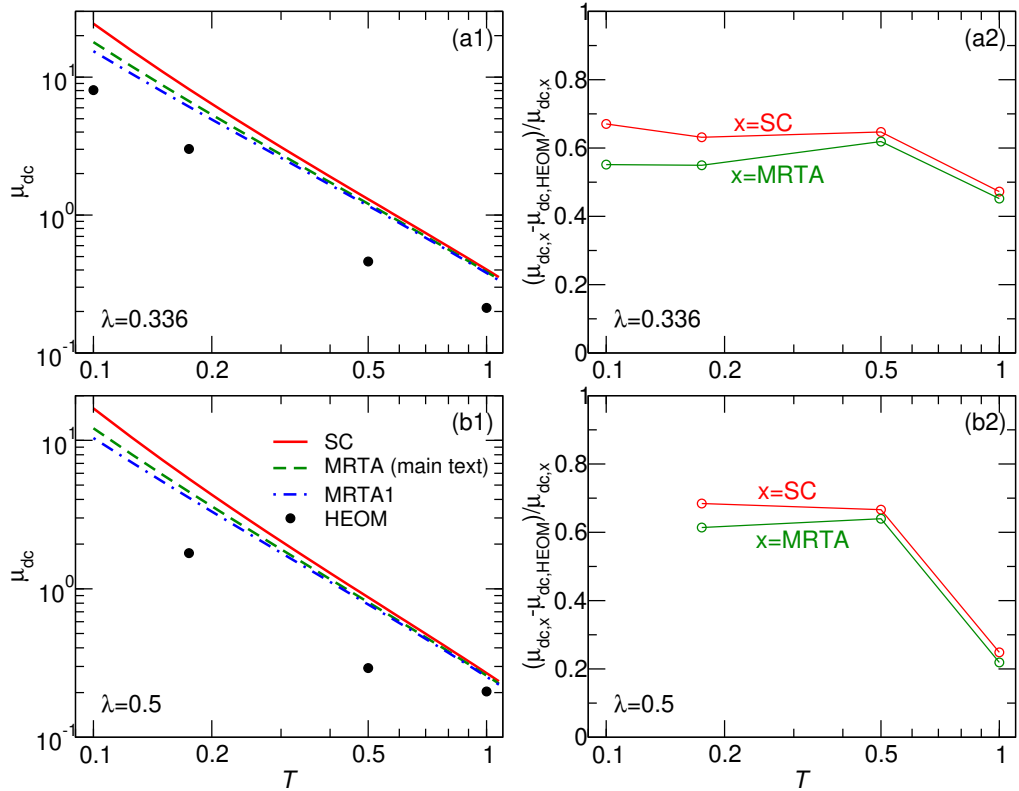}
    \caption{(a1) and (b1): Temperature-dependent carrier mobility computed within the HEOM approach (full symbols) and the Boltzmann transport theory with $\tau_k$ determined as a self-consistent solution to Eq.~\eqref{Eq:sc_tau_k_implicit} (label "SC", solid lines), in the MRTA [Eq.~\eqref{Eq:tau_k_MRTA}], dashed lines], and in the MRTA [Eq.~\eqref{Eq:tau_k_MRTA1}, dash-dotted lines].
    (a2) and (b2): Difference between the Boltzmann mobility and the HEOM mobility normalized to the Boltzmann mobility.
    In all panels, $\omega_0=0.044$ and $J=1$.
    The data labeled "SC" are the courtesy of N. Vukmirovi\'c.
    }
    \label{Fig:boltzmann_w0_0.044_201224}
\end{figure}

Figure~\ref{Fig:boltzmann_w0_0.044_201224} shows that the MRTA and MRTA1 mobilities are almost identical and close to those obtained in the self-consistent algorithm.
Therefore, the large differences between the HEOM and Boltzmann mobilities observed in the main text are not an artifact of the MRTA.

\newpage

\section{Comparison between HEOM results and TLS predictions for $\omega_0/J=0.5$}
\begin{figure}[htbp!]
    \centering
    \includegraphics[width=\textwidth]{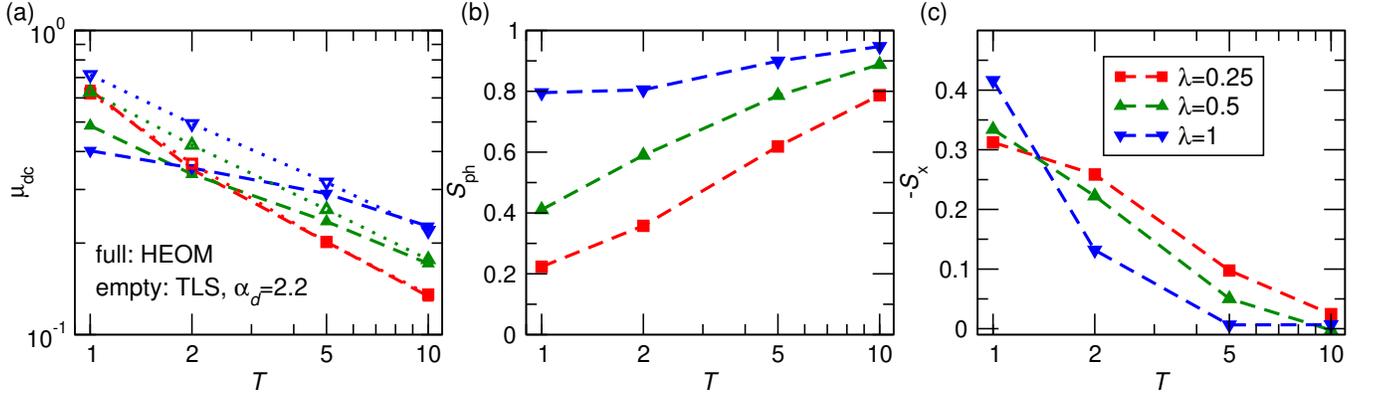}
    \caption{Temperature dependence of (a) the carrier mobility $\mu_\mathrm{dc}$, (b) the phonon-assisted share $S_\mathrm{ph}$ of $\mu_\mathrm{dc}$, and (c) the magnitude $-S_\mathrm{x}$ of the cross share of $\mu_\mathrm{dc}$ computed using HEOM (full symbols connected with dashed lines) and TLS with $\alpha_d=2.2$ (empty symbols connected with dotted lines).
    In all panels, we set $J=1$ and $\omega_0=0.5$.
    }
    \label{Fig:mu_vs_T_omega_0.5_251224}
\end{figure}

\begin{figure}[htbp!]
    \centering
    \includegraphics[width=0.5\textwidth]{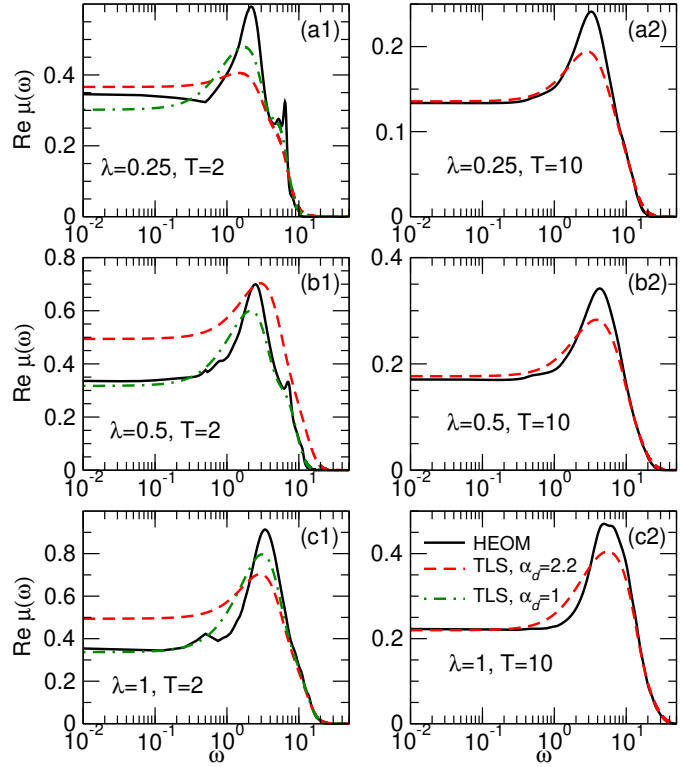}
    \caption{Dynamical-mobility profiles computed using HEOM (solid lines) and TLS [dashed lines and dashed-dotted lines in (a1)--(c1)] for different carrier--phonon interaction strengths at temperatures $T=2$ [(a1)--(c1)] and $T=10$ [(a2)--(c2)].
    The TLS results shown in panels (a1)--(c1) differ by the choice of the free parameter $\alpha_d$; the dashed and dashed-dotted lines display the results obtained using $\alpha_d=2.2$ and 1, respectively.}
    \label{Fig:optical_conductivities_w0_0.500_251224}
\end{figure}

In the main text, we suggest that the TLS can provide qualitative (and possibly even quantitative) insights into transport dynamics even when the timescales of free-carrier and free-phonon dynamics are comparable to one another.
Here, we compare and contrast HEOM results and TLS predictions for the temperature-dependent carrier mobility [Figs.~\ref{Fig:mu_vs_T_omega_0.5_251224}(a)--\ref{Fig:mu_vs_T_omega_0.5_251224}(c)] and the dynamical-mobility profile [Figs.~\ref{Fig:optical_conductivities_w0_0.500_251224}(a1)--\ref{Fig:optical_conductivities_w0_0.500_251224}(c2)] for the intermediate adiabaticity ratio $\omega_0/J=0.5$.

Comparing Fig.~\ref{Fig:optical_conductivities_w0_0.500_251224}(b) to Fig.~1(b) of the main text, we find that entering the regime of predominantly phonon-assisted transport with increasing $T$ or $\lambda$ is more rapid for $\omega_0/J=0.5$ than for $\omega_0/J=0.044$, as stated in the main text and concluded in the companion paper~\cite{part1}.
For example, fixing $\lambda=0.5$, we find $S_\mathrm{ph}\approx 0.8$ and $|S_\mathrm{x}|\approx 0.05$ for $\omega_0/J=0.5$ and $T/J=5$ ($T/\omega_0=10$), while $S_\mathrm{ph}\approx 0.3$ and $|S_\mathrm{x}|\approx 0.4$ for $\omega_0/J=0.044$ and $T/J=0.5$ (which is a somewhat higher temperature in units of $\omega_0$, $T/\omega_0\approx 12$).
Figure~\ref{Fig:mu_vs_T_omega_0.5_251224}(a) shows that the TLS (with $\alpha_d=2.2$) approximates HEOM mobilities very well at $T/J=10$.
The appropriateness of the TLS (with $\alpha_d=2.2$) at such high temperatures is further corroborated in Figs.~\ref{Fig:optical_conductivities_w0_0.500_251224}(a2)--\ref{Fig:optical_conductivities_w0_0.500_251224}(c2), which show that the TLS (with $\alpha_d=2.2$) reproduces the whole dynamical-mobility profile very well.
Figure~\ref{Fig:mu_vs_T_omega_0.5_251224}(a) also shows that the quality of TLS mobilities (with $\alpha_d=2.2$) generally deteriorates with decreasing temperature.
Meanwhile, Figs.~\ref{Fig:mu_vs_T_omega_0.5_251224}(b) and~\ref{Fig:mu_vs_T_omega_0.5_251224}(c) suggest that the TLS provides an overall correct physical picture even at $T/J=2$ and for $\lambda=0.5$ and 1, when the phonon-assisted contribution dominates over the band contribution ($S_\mathrm{ph}\gtrsim 0.5$), while the cross contribution is relatively small ($|S_\mathrm{x}|\lesssim \frac{1}{2}S_\mathrm{ph}$).
Indeed, by changing $\alpha_d$ from 2.2 to unity, we observe a much better agreement between the HEOM and TLS dynamical-mobility profiles in Figs.~\ref{Fig:optical_conductivities_w0_0.500_251224}(b1) and~\ref{Fig:optical_conductivities_w0_0.500_251224}(c1).

We believe that the very good performance of the TLS (with $\alpha_d=2.2$) for $\lambda=0.25$ throughout the temperature interval considered is somewhat fortuitous because we found it quite difficult to converge HEOM results with respect to $N$ and $D$ for $T/J=1$ and 2.
This is clearly reflected in relatively large (unsatisfactory) values of $\delta_\mathrm{OSR}$ in Table~\ref{Tab:details_w0_0.5} at lower temperatures and for moderate interactions.
Although changing $\alpha_d$ from 2.2 to unity yields an overall better agreement between HEOM and TLS dynamical-mobility profiles for $\lambda=0.25$ and $T/J=2$, see Fig.~\ref{Fig:optical_conductivities_w0_0.500_251224}(a1), the values of $S_\mathrm{ph}$ and $|S_\mathrm{x}|$ suggest that the details of the physical picture are somewhat different from those characteristic of the TLS.
Indeed, the HEOM dynamical-mobility profile in Fig.~\ref{Fig:optical_conductivities_w0_0.500_251224}(a1) displays a weakly pronounced dip at $\omega\approx\omega_0$ and a weakly pronounced rise in $\mathrm{Re}\:\mu(\omega)$ as $\omega\to 0$.

\newpage

\section{HEOM results for $\omega_0/J=0.044,\lambda=0.5,$ and $T/J=0.175$. Comparison with the results of quantum--classical simulations}
\label{App:quantum-classical}
\begin{figure}[htbp!]
    \centering
    \includegraphics[width=\textwidth]{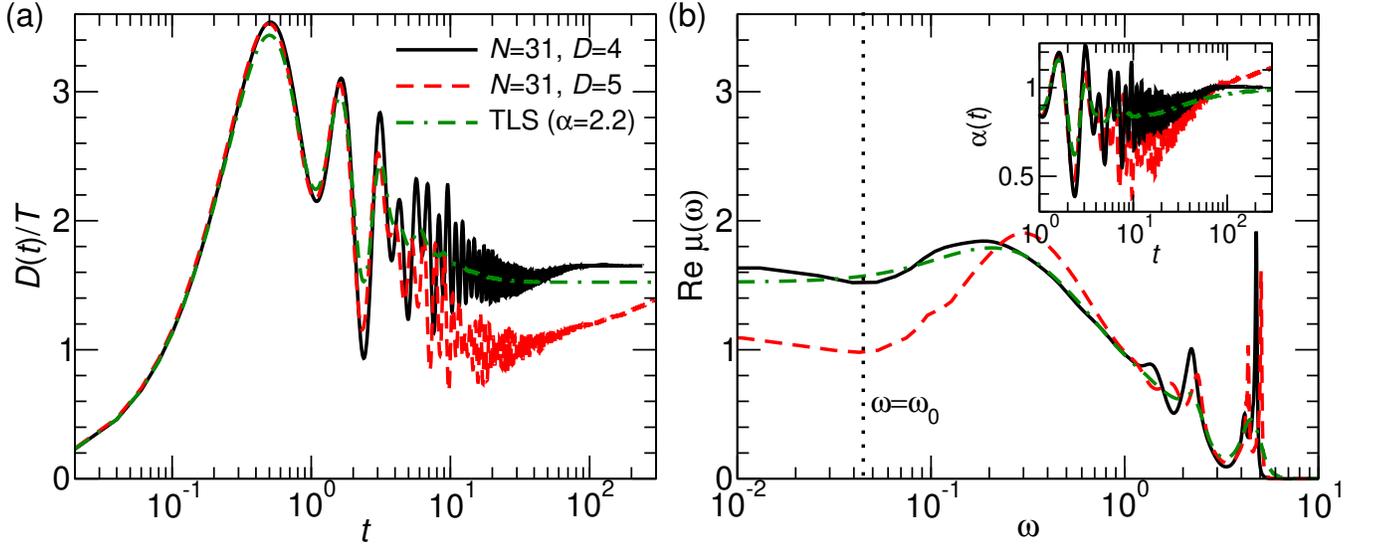}
    \caption{Comparison of (a) the quantity $\mathcal{D}(t)/T$ and (b) the dynamical-mobility profile computed by solving the HEOM for $N=31,D=4$ and $N=31,D=5$ and assuming the TLS with $\alpha_d=2.2$. We plot $\mathcal{D}(t)/T$ instead of $\mathcal{D}(t)$ to enable a direct comparison of the long-time dynamics in (a) and low-frequency spectrum in (b).
    The inset in (b) shows the dynamics of the diffusion exponent.
    The model parameters are set to $J=1,\omega_0=0.044,\lambda=0.5,$ and $T=0.175$, so that a direct comparison with the room-temperature quantum--classical simulations of Ref.~\onlinecite{runeson2024} can be performed.}
    \label{Fig:result_mash}
\end{figure}
Figures~\ref{Fig:result_mash}(a) and~\ref{Fig:result_mash}(b) present HEOM results in the parameter regime that has been recently studied using the quantum--classical simulations of Ref.~\onlinecite{runeson2024}, and compare them with the TLS predictions.
The authors of Ref.~\onlinecite{runeson2024} employ a different definition of the dimensionless interaction parameter $\lambda$, so that their value of 0.25 corresponds to our value of 0.5.
They also use a different value of the transfer integral, $J=110\:\mathrm{meV}$, while we use $J=143\:\mathrm{meV}$.
Our temperature $T/J=0.175$ then corresponds to the physical temperature $T=291\:\mathrm{K}$, so that our HEOM results should be compared with the quantum--classical results at $T=300\:\mathrm{K}$.

A comparison of Fig.~\ref{Fig:result_mash}(a) with Fig.~2(a) of Ref.~\onlinecite{runeson2024} suggests that the quantum--classical simulations qualitatively reproduce the dynamics of the numerically exact diffusion constant.
The decrease in the quantum--classical $\mathcal{D}(t)$ on intermediate timescales seems to be more consistent with the HEOM results for $D=5$ than with the results for $D=4$.
Also, the position of the finite-frequency peak for $N=31,D=5$ ($\omega_\mathrm{DDP}/J\approx 0.3$ or $\hbar\omega_\mathrm{DDP}\approx 345\:\mathrm{cm}^{-1}$ in physical units) seems to be more consistent with the $\approx 400\:\mathrm{cm}^{-1}$ DDP reported in Fig.~3(a) of Ref.~\onlinecite{runeson2024} than the position of the finite-frequency peak for $N=31,D=4$ ($\omega_\mathrm{DDP}/J\approx 0.18$ or $\hbar\omega_\mathrm{DDP}\approx 210\:\mathrm{cm}^{-1}$ in physical units).
However, we prefer HEOM results for $D=4$ to those for $D=5$ because the latter enable us to reliably estimate $\mu_\mathrm{dc}$.
The dynamics of the diffusion exponent $\alpha$ in the inset of Fig.~\ref{Fig:result_mash}(b) shows that HEOM results for $D=4$ do approach the long-time diffusive regime on the timescales we consider, while the diffusive transport is not apparent in HEOM results for $D=5$.
The diffusive transport sets in from the superdiffusive side, reflecting the long-time increase in the diffusion constant.
The overshoot of $\alpha(t)$ above unity is somewhat less pronounced than in Fig.~2(a1) of the main text, which is consistent with the increase in the interaction constant from $\lambda=0.336$ to 0.5.

The quantum--classical carrier mobility for $J=110\:\mathrm{meV},\omega_0=0.0435\:J,$ and $T=300\:\mathrm{K}$ is $\mu_\mathrm{dc}^\mathrm{QC}\approx 8\:\mathrm{cm}^2/\mathrm{(Vs)}$, so that the value that should be compared with our HEOM result $\mu_\mathrm{dc}^\mathrm{HEOM}\approx 13\:\mathrm{cm}^2/(\mathrm{Vs})$ is $\frac{143}{110}\times\frac{300}{291}\mu_\mathrm{dc}^\mathrm{QC}\approx 11\:\mathrm{cm}^2/(\mathrm{Vs})$.
Therefore, the quantum--classical carrier mobility agrees reasonably well with the HEOM mobility for $N=31,D=4$.
Interestingly, the TLS mobility also agrees well with the HEOM mobility for $N=31,D=4$, while the TLS dynamical-mobility profile in Fig.~\ref{Fig:result_mash}(b) appears to be more consistent with HEOM results for $D=4$ than for $D=5$.

\bibliography{aipsamp}